%% file: main.tex
\documentclass{article}

\PassOptionsToPackage{numbers, compress}{natbib}
\usepackage[preprint]{neurips_2026}
\usepackage[utf8]{inputenc}
\usepackage[T1]{fontenc}
\usepackage{url}
\usepackage{booktabs}
\usepackage{amsfonts}
\usepackage{nicefrac}
\usepackage{microtype}
\usepackage{xcolor}
\usepackage{colortbl}
\definecolor{rankgold}{HTML}{FFE38A}
\definecolor{ranksilver}{HTML}{D7E5F5}
\usepackage{graphicx}
\usepackage{amsmath}
\usepackage{amssymb}
\usepackage{enumitem}
\usepackage{multicol}
\usepackage{array}
\usepackage{tabularx}
\usepackage{xspace}
\usepackage{float}
\usepackage{makecell}
\usepackage{multirow}
\usepackage{tikz}
\usepackage{wrapfig}
\usepackage{algorithm}
\usepackage{algpseudocode}
\usetikzlibrary{positioning,arrows.meta,calc,fit,backgrounds,shapes.geometric}
\newcolumntype{Y}{>{\raggedright\arraybackslash}X}
\newcolumntype{L}[1]{>{\raggedright\arraybackslash}p{#1}}
\newcolumntype{C}[1]{>{\centering\arraybackslash}p{#1}}
\usepackage{ragged2e}
\usepackage{placeins}

\usepackage[table]{xcolor}
\definecolor{lightgray}{gray}{0.92}

\usepackage{xcolor}

\newcommand{\benchmarkfullname}{MultiMedia-TerminalBench\xspace}
\newcommand{\benchmarkabbr}{MMTB\xspace}
\newcommand{\benchmark}{\benchmarkabbr}

\newcommand{\grace}{Terminus-MM\xspace}
\newcommand{\kira}{Terminus-KIRA\xspace}
\newcommand{\rocky}{Terminus-A\xspace}

\newcommand{\termii}{Terminus-2\xspace}

\definecolor{myblue}{rgb}{0.12,0.35,0.62}
\usepackage[pagebackref,breaklinks,colorlinks,allcolors=myblue]{hyperref}

\usepackage{pifont}
\usepackage{wasysym}
\usepackage{caption}
\newcolumntype{Y}{>{\centering\arraybackslash}X}
\newcolumntype{L}[1]{>{\raggedright\arraybackslash}p{#1}}
\newcommand{\cmark}{\textcolor{green!50!black}{\scalebox{0.9}{\CIRCLE}}}
\newcommand{\xmark}{\textcolor{red}{\ding{55}}}
\newcommand{\tmark}{\textcolor{orange!90!black}{\scalebox{1.15}{$\blacktriangle$}}}

\usepackage{pgfplots}
\pgfplotsset{compat=1.18}
\definecolor{mpblue}{RGB}{52,102,187}
\definecolor{mporange}{RGB}{230,140,20}
\definecolor{mpred}{RGB}{210,70,60}
\definecolor{mppurple}{RGB}{120,82,180}
\definecolor{mpgreen}{RGB}{90,165,70}

\title{MMTB: Evaluating Terminal Agents \\ on Multimedia-File Tasks}

\author{%
  \textbf{Chiyeong Heo$^{1}$, Jaechang Kim$^{1}$, Junhyuk Kwon$^{1}$, Hoyoung Kim$^{3}$} \\[2pt]
  \textbf{Dongmin Park$^{4}$, Jonghyun Lee$^{4}$, Jungseul Ok$^{1,2}$\thanks{Corresponding author.}} \\[2pt]
  $^{1}$GSAI, POSTECH \quad $^{2}$CSE, POSTECH \quad $^{3}$National AI Research Lab \quad $^{4}$Krafton AI \\[2pt]
  \url{https://mm-tbench.github.io/multimedia-terminal-bench/}
}

\begin{document}

\maketitle

\input{01abstract}

\input{02intro}

\input{03benchmark_and_tasks}
\input{04evaluation_setup}

\input{05main_results}

\input{07related_work}
\input{08discussion}

% Body restructure 2026-04-29: previous body files (03benchmark_setting,
% 04design_principles, 05task_suite, 06evaluation_protocol, 07baseline_methods,
% 08results, 09analysis, 10related_work, 11discussion, 12conclusion) are kept
% as orphans for diff history; the new \input chain above replaces them.

\section{Conclusion}
We introduced MMTB, a rigorously validated benchmark of 105 realistic multimedia-file tasks in persistent terminal workspaces, together with Terminus-MM, a workspace-aware harness that enables native access to audio and video files. MMTB is grounded in practical practitioner workflows, packaged as self-contained Harbor tasks, and filtered through automated checks, oracle solvability tests, baseline screening, and manual validation. Our experiments show that native multimedia access improves agents’ ability to complete these practical workflows, while terminal-only approaches often rely on longer and costlier command-line evidence-gathering pipelines. These results highlight the need for future multimedia terminal agents that combine direct audio-visual grounding with reliable shell execution and artifact construction.

% We introduce \benchmark{}, a 105-task benchmark for evaluating terminal agents on multimedia-file tasks, together with Terminus-MM, a harness that extends Terminus-KIRA with native audio and video perception. Our experiments show that native multimedia access helps agents solve tasks requiring evidence from audio and video files, while agents without such access rely on additional command-line workflows that increase trajectory cost. The non-nested successes of Terminus-MM and Codex CLI further show that multimedia perception and shell execution address complementary parts of the benchmark. Future multimedia terminal agents should therefore combine joint audio-visual reasoning with reliable shell scripting for artifact construction.

% We introduce \benchmark{}, a 105-task benchmark for terminal agents on
% multimedia-file workflows, and Terminus-MM, a controlled harness
% that extends Terminus-KIRA with native audio and video perception. Across
% four backbones and seven Terminus variants, native multimedia access
% improves over text-only and image-only harnesses on audio- and
% video-grounded tasks, and shell-mediated proxy reconstruction (ASR
% transcripts, sampled frames, signal-processing scripts) closes only part
% of the gap while substantially inflating trajectory cost.
% Terminus-MM and Codex CLI succeed on overlapping but non-nested
% task sets, indicating that future multimedia terminal agents will need
% both joint audio-visual reasoning and robust shell scripting in a single
% harness.

% \paragraph{Acknowledgments}
% Keep acknowledgments only for non-anonymous versions.

\bibliographystyle{plainnat}
\bibliography{references}

\newpage
\appendix

\input{09appendix}

% Uncomment if you want to include the official checklist in the source.
% arXiv preprint: omit reviewer checklist.
% \input{checklist}

\end{document}

%% file: 01abstract.tex
\begin{abstract}
Terminals provide a powerful interface for AI agents by exposing diverse tools for automating complex workflows, yet existing terminal-agent benchmarks largely focus on tasks grounded in text, code, and structured files. However, many real-world workflows require practitioners to work directly with audio and video files. Working with such multimedia files calls for terminal agents not only to understand multimedia content, but also to convert auditory and visual evidence across related files into appropriate actions. To evaluate terminal agents on multimedia-file tasks, we introduce MultiMedia-TerminalBench (MMTB), a benchmark of 105 tasks across 5 meta-categories where terminal agents directly operate with audio and video files. Alongside MMTB, we propose Terminus-MM, a multimedia harness that extends Terminus-KIRA with audio and video perception for terminal agents. Together, MMTB and Terminus-MM support a controlled study of multimedia terminal agents, revealing how different forms of multimedia access shape task outcomes and determine which evidence agents rely on to construct executable terminal workflows. MMTB media and metadata are released at \url{https://huggingface.co/datasets/mm-tbench/mmtb-media}.
\end{abstract}

%% file: 02intro.tex
% figure 1
\input{tex-table-figure/qualitative_main_figure}
% figure 1

\input{tex-table-figure/benchmark_comparison_table}

\section{Introduction}
\label{sec:intro}
% \blfootnote{Code mirror: \url{https://github.com/mm-tbench/multimedia-terminal-bench}.}
As terminals provide a powerful interface for AI agents, recent terminal agents such as Claude Code~\cite{anthropic2026claudecode} and Codex CLI~\cite{openai2026codexcli} have emerged as practical tools for automating complex command-line workflows. By utilizing shell commands and external tools, terminal agents can interact with files, execute code, search the web, and generate outputs in persistent workspaces. Such capabilities make terminals a natural environment for evaluating realistic workflows, from planning and tool use to verifiable task completion. Accordingly, recent terminal-agent benchmarks such as Terminal-Bench~\cite{merrill2026terminalbench} evaluate terminal agents on diverse realistic workflows using the Harbor task format, where each task consists of a user instruction, a working directory, and an expected output specification.

Despite this progress, current terminal-agent benchmarks focus primarily on tasks grounded in text, code, and structured files~\cite{merrill2026terminalbench,yang2023intercode}. However, many real-world workflows require practitioners to work directly with multimedia files such as audio and video recordings. For instance, users may need to prepare media for broadcast or social platforms~\cite{qiu2024mmsum}, provide feedback on music or acting performances~\cite{ashutosh2025expertaf}, process meetings or compliance-sensitive recordings~\cite{hu2023meetingbank}, or annotate audio-visual data for research~\cite{gemmeke2017audio}. 
Supporting such workflows requires terminal agents to move beyond multimedia understanding alone. They must ground decisions in auditory and visual evidence across files and execute the corresponding actions in a terminal environment. However, existing benchmarks lack multimedia-file tasks designed to evaluate terminal agents~\cite{jang2024videowebarena,li2024omnibench,merrill2026terminalbench}.

To this end, we introduce MultiMedia-TerminalBench (MMTB), a benchmark centered on multimedia-file tasks in terminals. 
As shown in Table~\ref{tab:comp_bench}, \benchmark{} differs from prior computer-use benchmarks, including Terminal-Bench~\cite{merrill2026terminalbench}, which provide limited coverage of audio and video files and content-aware reasoning over them. 
It also differs from audio-visual benchmarks focused mainly on multimedia understanding~\cite{bie2025omniplay,chao2025jointavbench,chowdhury2025avtrustbench,li2024omnibench}.
Specifically, MMTB consists of 105 tasks across 5 meta-categories, with each task grounded in a public source reflecting a paid practitioner workflow, such as those on Upwork or Fiverr websites, and packaged in the Harbor format used by Terminal-Bench~\cite{merrill2026terminalbench}.
% , and packaged in the Harbor format~\cite{merrill2026terminalbench}.
% Specifically, MMTB consists of 105 tasks across 5 meta-categories, \textcolor{blue}{each anchored to a specific public URL documenting a paid practitioner workflow on Upwork, Fiverr, casting and voiceover platforms, or industry forums and standards}, with each task following the Harbor format~\cite{merrill2026terminalbench}.
Figure~\ref{fig:overview} further illustrates this design through a workspace example containing a task instruction and the corresponding audio and video files.

To solve the example in Figure~\ref{fig:overview}, conventional terminal agents such as Terminus-2~\cite{merrill2026terminalbench} and Codex CLI~\cite{openai2026codexcli} rely on the intermediate representations of the given multimedia files, rather than directly perceiving their audio or video content. For example, audio may be transformed into spectrograms or RMS signals, while video may be reduced to extracted frames. These additional processing steps can increase the required time and discard important information during conversion. Quantitatively, we observe that the standalone terminal agent baseline, Codex CLI with GPT-5.2, solves only $16.2\%$ of MMTB tasks, revealing the limitations of conventional terminal agents on multimedia-file tasks.

% \textcolor{blue}{cannot read audio or video files directly: they have to convert each file into text or numeric features through command-line tools --- a transcript produced by speech recognition, a handful of frames decoded from a video, or summary statistics computed from an audio waveform --- and then reason over those derived numbers and strings. Each conversion adds processing steps and discards information about the original media.} 

To address these limitations, we introduce \textsc{Terminus-MM}, a multimedia terminal-agent harness that extends Terminus-KIRA~\cite{kira2024} with audio and video perception. In addition, \textsc{Terminus-MM} adapts its perception interface to each workspace by exposing tools matched to the available multimedia files. Using this workspace-aware design, we compare audio-only, video-only, and combined audio-video access to analyze how different forms of multimedia perception affect task outcomes and which observed evidence routes agents rely on when constructing executable terminal artifacts.

Our main contributions are summarized as follows:
\begin{enumerate}[leftmargin=1.4em]
\item We introduce \benchmark{}, a benchmark for evaluating terminal agents on multimedia-file tasks, where terminal agents inspect diverse multimedia files, ground terminal actions in multimedia evidence, and produce verifiable output artifacts.
% \item We introduce \benchmark{} to evalaute terminal agents on multimedia-file tasks: persistent file-based workflows in which agents must inspect diverse multimedia files, ground terminal actions in multimedia evidence, and produce verifiable output artifacts. 
% \benchmark{} is built on the Harbor task format with artifact-level evaluation and reproducible terminal execution.

\item Alongside \benchmark{}, we propose \textsc{Terminus-MM}, a multimedia terminal-agent harness extending Terminus-KIRA with audio and video perception, whose interface is adapted to the multimedia files available in each workspace.
% \item We propose \textsc{Terminus-MM}, a model-agnostic shell-centered
% harness for evaluating omni models on persistent multimedia workspaces. It
% combines terminal access with deterministic workspace-routed native image,
% audio, and video perception tools, and its modality-restricted variants support
% controlled access ablations.
% \item We provide a shared shell-centered evaluation protocol for multimedia terminal agents, including \grace{}, a harness that extends \kira{} with audio, and video perception. This protocol enables controlled modality comparisons across text-only, text+image, text+audio/video, and full-multimedia agents while keeping the task interface fixed.

\item We analyze the tasks solved by different terminal agents on \benchmark{} to reveal how multimedia access shapes task outcomes and evidence use in executable terminal workflows.
% \item We show that \benchmark{} remains far from saturated by current omni
% and terminal agents, with the best full-multimedia setting solving fewer
% than half of the tasks. We further diagnose this gap using modality
% ablations, \textsc{Terminus-MM}+Gemini-3.1-Pro versus Codex CLI+GPT-5.2
% task overlap, proxy-cost measurements, failure traces, and tool-routing
% ablations.
\end{enumerate}

%% file: tex-table-figure/qualitative_main_figure.tex
\begin{figure}[H]
\centering
\includegraphics[width=\linewidth, trim={0.8cm 0 0.8cm 0},]{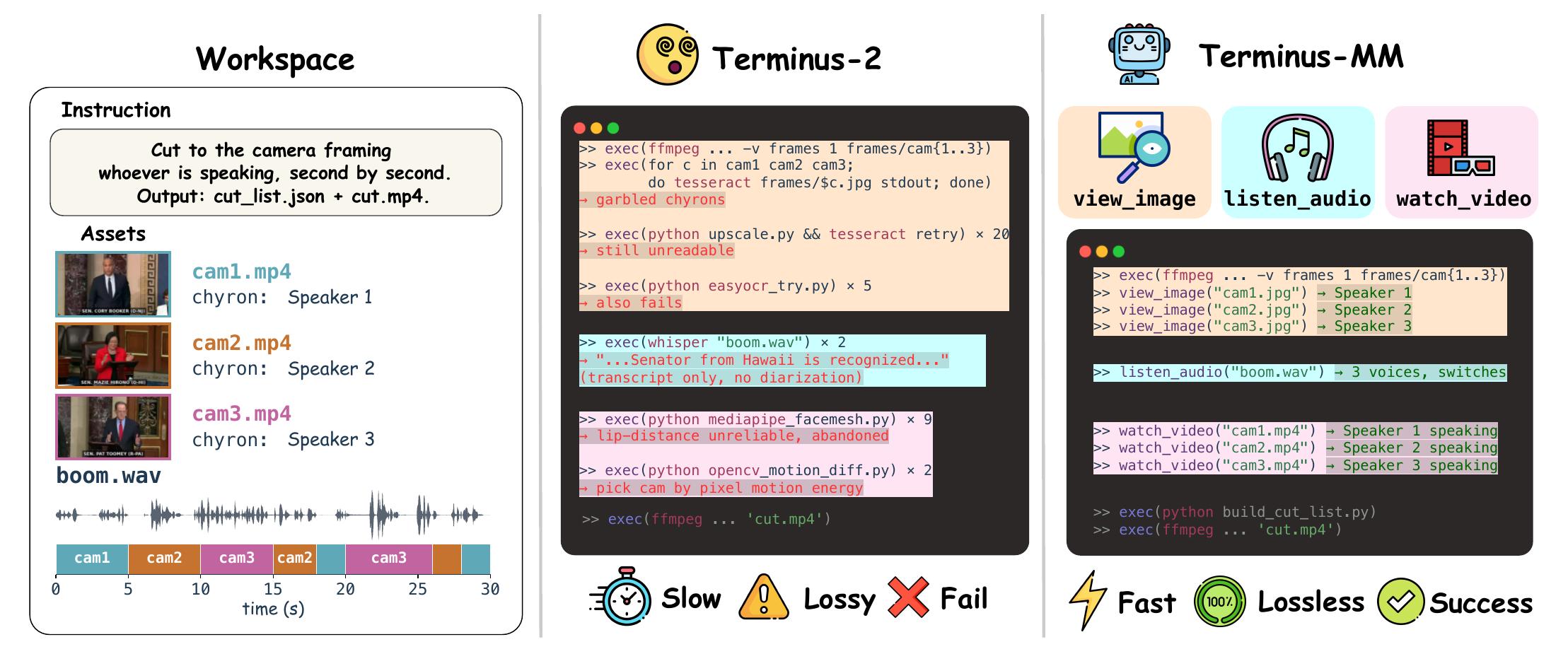}

\caption{
\textbf{An example \benchmark{} task and two terminal-agent approaches.}
The task merges three videos and one audio file into one edited
artifact. Agents with native multimodal access read the raw files
directly; text-only agents must reach the same evidence through
command-line tools (OCR, ASR, motion-energy), adding processing steps
that introduce inefficiency and errors.
}

\label{fig:overview}
\end{figure}

\iffalse
\caption{Overview of \benchmark{}. On the
\texttt{multicam-active-speaker-cut} task, \textsc{Terminus-MM} reads
chyrons (\texttt{view\_image}), separates voices in the boom mix
(\texttt{listen\_audio}), and aligns the active-speaker timeline
(\texttt{watch\_video}) before assembling \texttt{cut\_list.json} and
\texttt{cut.mp4}. \textsc{Terminus-2}, lacking native perception,
reconstructs each modality through proxy pipelines (OCR, ASR, motion
energy); each proxy loses information and the agent ends up picking
cameras by raw pixel motion alone.
}
\fi

%% file: tex-table-figure/benchmark_comparison_table.tex
\newcolumntype{C}[1]{>{\centering\arraybackslash}p{#1}}

\begin{table*}[t]
\centering
\small
\vspace{-4mm}
\caption{
\textbf{Comparison of \benchmark{} with existing computer-use and audio-visual benchmarks.}
\textit{T}, \textit{I}, \textit{A}, and \textit{V} denote text, image, audio, and video, respectively.
\tmark denotes partial coverage: the benchmark addresses the cross-file aspect but not the audio-visual aspect.
%
\iffalse
\textbf{Comparison of \benchmark{} with existing audio-visual and computer-use benchmarks.}
\textit{T}, \textit{I}, \textit{A}, and \textit{V} denote text, image, audio, and video, respectively.
Parentheses indicate incidental media artifacts rather than core benchmark modalities. \tmark denotes partial coverage of the column's capability — the benchmark addresses the cross-file or workflow aspect but not the audio-visual reasoning component (e.g., cross-file reasoning over text or code).
% Parentheses indicate incidental media artifacts rather than a core benchmark modality.
% \tmark indicates partial coverage, implying does not centrally require content-aware reasoning over multiple media files.
% the benchmark may involve file or environment state, but does 
\fi
}
\vspace{-1mm}
\label{tab:comp_bench}
% \begin{tabularx}{0.8\linewidth}{L{3.15cm}L{1.25cm}YYY}
\begin{tabularx}{0.865\linewidth}{L{3.0cm}L{1.15cm}C{2cm}C{2cm}C{2cm}}
\toprule
\multicolumn{1}{l}{\multirow{2}{*}{Benchmarks}} &
\multirow{2}{*}{Input} &
Content-aware &
Cross-file &
Persistent \\
& & AV reasoning & AV reasoning & file workflow \\
\midrule

InterCode~\citep{yang2023intercode}
& $T$ & \xmark & \tmark & \cmark \\

OSWorld~\citep{xie2024osworld}
& $T{+}I$ & \xmark & \tmark & \cmark \\

Terminal-Bench~\citep{merrill2026terminalbench}
& $T$ & \xmark & \tmark & \cmark \\

\midrule

OmniBench~\citep{li2024omnibench}
& $T{+}I{+}A$ & \cmark & \xmark & \xmark \\

JointAVBench~\citep{chao2025jointavbench}
& $T{+}A{+}V$ & \cmark & \xmark & \xmark \\

AVTrustBench~\citep{chowdhury2025avtrustbench}
& $T{+}A{+}V$ & \cmark & \xmark & \xmark \\

OmniPlay~\citep{bie2025omniplay}
& $T{+}I{+}A{+}V$ & \cmark & \xmark & \xmark \\

\midrule

VideoWebArena~\citep{jang2024videowebarena}
& $T{+}I{+}A{+}V$ & \cmark & \xmark & \xmark \\

\benchmark{} (ours)
& $T{+}I{+}A{+}V$ & \cmark & \cmark & \cmark \\

\bottomrule
\end{tabularx}

% \vspace{-4mm}
\end{table*}

%% file: 03benchmark_and_tasks.tex
\input{tex-table-figure/benchmark_construction_figure}

\section{\benchmarkfullname: Benchmark Design}
\label{sec:benchmark-design}

In this section, we describe the design of \benchmark{} in three parts, moving from the benchmark scope, to the construction pipeline for building and filtering tasks, and finally to the task organization and Harbor-based format that makes individual tasks self-contained and reusable across agents.

\paragraph{Benchmark scope.}
\benchmark{} evaluates terminal agents on tasks where multimedia files are the
central objects of work, with a particular focus on audio and video files. Each
task takes place in a persistent terminal workspace containing multimedia assets
that the agent must inspect, edit, or convert into a verifiable
output. Unlike existing multimedia question-answering benchmarks, where multimedia files
typically provide context for answering questions in text, \benchmark{} requires
agents to use multimedia evidence to execute terminal actions and produce the
required artifact.
Existing terminal-agent benchmarks center on text artifacts:
Terminal-Bench's 80 tasks, for example, cover software engineering, sysadmin,
and data analysis but contain no multimedia inputs. 

\paragraph{Benchmark construction pipeline.}
Figure~\hyperref[fig:mmtb-construction-coverage]{\ref*{fig:mmtb-construction-coverage}a} summarizes the construction pipeline for \benchmark{}. We begin with 163 candidate scenarios, each anchored to a specific public URL documenting a paid practitioner workflow --- predominantly Upwork and Fiverr gig listings,\footnote{\url{https://www.upwork.com/}; \url{https://www.fiverr.com/}.} with casting calls, practitioner forums, and industry standards documents making up the rest --- so the suite captures the multimedia work people are actually paid to do, not synthetic instruction templates.
Each scenario is scoped into a concrete task design, scaffolded as a Harbor task, and populated with license-compatible external media or controlled synthetic and derivative assets. The resulting candidate tasks are then filtered through automated checks for task structure, Docker build, media fetching, oracle solvability, and dummy/no-op failure, as well as baseline checks for tasks easily solved by baselines, followed by manual review for trivial shortcuts, unrealistic setups, and redundancy.
After filtering, we obtain 105 tasks, each with asset provenance recorded in \texttt{media.toml}, including source descriptions, license information, and content hashes.
Additional details are provided in Appendices~\ref{sec:appendix-release} and~\ref{sec:appendix-source-corpus}.

\paragraph{Task categories and task format.}
The final 105 tasks are organized into 5 meta-categories covering practical multimedia workflows, including media production, performance and coaching, enterprise and compliance, personal and education, and operations and research, as shown in Figure~\hyperref[fig:mmtb-construction-coverage]{\ref*{fig:mmtb-construction-coverage}b}. Within these meta-categories, tasks are further divided into 16 fine-grained workflow categories. We also annotate each task with multi-label capability tags, summarized in Figure~\hyperref[fig:mmtb-construction-coverage]{\ref*{fig:mmtb-construction-coverage}c}, which capture the perceptual and reasoning capabilities required by the task. Since these tags are not mutually exclusive, their counts are marginal frequencies rather than a partition of the 105 tasks. Frequent tag co-occurrences are reported in Figure~\ref{fig:tag-cooccurrence}.
The corpus totals 536 media files and 6\,h\,54\,min of timed audio-visual content, with a median per-task duration of 1\,m\,20\,s. A human practitioner could plausibly walk through the entire suite in a few hours of skimming and scrubbing. Detailed media-file statistics and an analysis of how agent performance varies with task duration are in Appendix~\ref{app:media-volume}.

Beyond this suite-level organization, each task is implemented as a Harbor task unit following Terminal-Bench~\citep{merrill2026terminalbench}, with five components: (i) an \emph{instruction}, which states the user's goal and required deliverable without exposing the answer; (ii) a \emph{workspace}, which provides a containerized filesystem with stored multimedia files and optional supporting files; (iii) an \emph{allowed terminal/tool interface}, which defines the operations exposed to the agent under the evaluation harness; (iv) an \emph{output schema}, which specifies the required artifact path and format; and (v) an \emph{artifact evaluator}, which scores the produced artifact. Following the Harbor protocol, each task is scored by evaluating the final artifact submitted at the required path, rather than the agent's rationale or command trace. The expected artifact varies by task and may take the form of a selected file, a timestamp or interval, a structured JSON/CSV record, an edit list, or a processed media file.

The Harbor format also makes MMTB accessible and agent-agnostic: each task is a self-contained unit that benchmark users can inspect and run without reconstructing task-specific setup or scoring logic. Since the same task unit can be evaluated by swapping only the agent while keeping the workspace and evaluator fixed, the format naturally supports comparisons across terminal agents.

%% file: tex-table-figure/benchmark_construction_figure.tex
\begin{figure*}[!t]
\centering
\small

% ============================================================
% (a) Task construction and release filtering
% ============================================================
\vspace{-2mm}
\includegraphics[width=\linewidth]{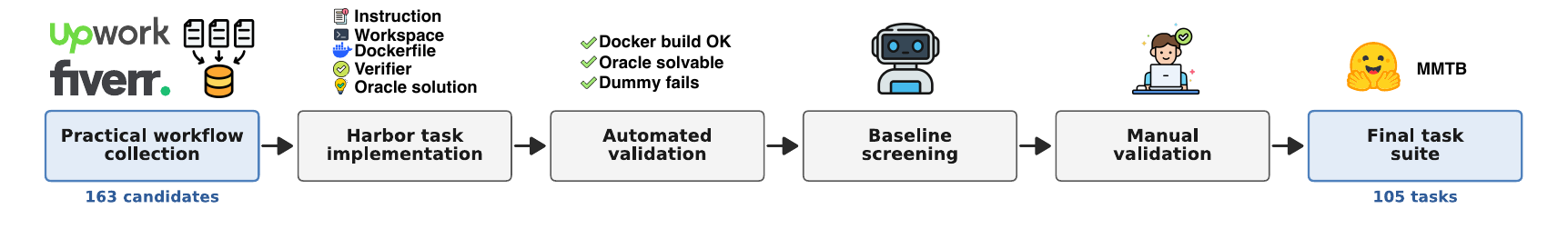}
\vspace{-1mm}
{\small (a) Benchmark Construction Pipeline}

\vspace{1em}

% ============================================================
% (b) Workflow taxonomy
% ============================================================
\hfill
\begin{minipage}[t]{0.50\linewidth}
\centering
\resizebox{\linewidth}{!}{%
\begin{tikzpicture}
\begin{axis}[
    width=2cm,
    height=4cm,
    scale only axis,
    xbar,
    xmin=0,
    xmax=20,
    ymin=-0.4,
    ymax=21,
    axis x line*=bottom,
    axis y line*=left,
    y axis line style={draw=none},
    xlabel={\# of tasks},
    xlabel style={font=\tiny},
    xtick={0,5,10,15,20},
    xticklabel style={font=\tiny},
    ytick={
        0.3,1.55,2.8,
        4.5,5.7,
        7.3,8.5,9.7,
        11.3,12.5,13.7,
        15.3,16.5,17.8,19.0,20.2
    },
    yticklabel pos=left,
    yticklabels={
        Public Safety,
        Automation,
        Dataset Annotation,
        Personal / Everyday,
        Education \& Lecture,
        Document Processing,
        Compliance \& Privacy,
        Corporate Meetings,
        Language Learning,
        Acting \& Casting,
        Music Coaching,
        Creator Economy,
        Audio Eng.\ \& Podcast,
        Game QA \& Esports,
        Subtitling \& Localization,
        Broadcast \& Film
    },
    yticklabel style={
        font=\tiny,
        anchor=east,
        align=right,
        xshift=1pt
    },
    ytick style={draw=none},
    xtick style={draw=none},
    xmajorgrids=true,
    ymajorgrids=false,
    grid style={dashed, gray!30},
    bar width=4.1pt,
    nodes near coords,
    every node near coord/.append style={
        font=\tiny,
        scale=0.8,
        anchor=west,
        xshift=1pt
    },
    clip=false
]

% Operations & Research
\addplot[draw=none, fill=mpgreen] coordinates {
    (1,2.5)
    (1,3.7)
    (4,4.9)
};

% Personal & Education
\addplot[draw=none, fill=mppurple] coordinates {
    (5,5.5)
    (8,6.7)
};

% Enterprise & Compliance
\addplot[draw=none, fill=mpred] coordinates {
    (4,7.3)
    (6,8.5)
    (13,9.7)
};

% Performance & Coaching
\addplot[draw=none, fill=mporange] coordinates {
    (6,10.3)
    (8,11.5)
    (9,12.7)
};

% Media Production
\addplot[draw=none, fill=mpblue] coordinates {
    (3,13.3)
    (6,14.5)
    (6,15.7)
    (8,16.9)
    (17,18.1)
};

% Group labels (left)
\node[anchor=east, font=\bfseries\tiny, text=mpblue, xshift=-1cm]
    at (rel axis cs:-0.70,0.84) {Media Production};
\node[anchor=east, font=\bfseries\tiny, text=mporange, xshift=-1cm, yshift=0.1cm]
    at (rel axis cs:-0.70,0.57) {Performance \& Coaching};
\node[anchor=east, font=\bfseries\tiny, text=mpred, xshift=-1cm, yshift=0.1cm]
    at (rel axis cs:-0.70,0.38) {Enterprise \& Compliance};
\node[anchor=east, font=\bfseries\tiny, text=mppurple, xshift=-1cm, yshift=0.12cm]
    at (rel axis cs:-0.70,0.22) {Personal \& Education};
\node[anchor=east, font=\bfseries\tiny, text=mpgreen, xshift=-1cm, yshift=0.05cm]
    at (rel axis cs:-0.70,0.08) {Operations \& Research};

% Group brackets (moved left like panel (c))
\draw[mpblue, thick, xshift=-1.7cm]
    (rel axis cs:-0.34,0.71) -- (rel axis cs:-0.37,0.71) --
    (rel axis cs:-0.37,0.985) -- (rel axis cs:-0.34,0.985);

\draw[mporange, thick, xshift=-1.7cm]
    (rel axis cs:-0.34,0.52) -- (rel axis cs:-0.37,0.52) --
    (rel axis cs:-0.37,0.69) -- (rel axis cs:-0.34,0.69);

\draw[mpred, thick, xshift=-1.7cm]
    (rel axis cs:-0.34,0.33) -- (rel axis cs:-0.37,0.33) --
    (rel axis cs:-0.37,0.50) -- (rel axis cs:-0.34,0.50);

\draw[mppurple, thick, xshift=-1.7cm]
    (rel axis cs:-0.34,0.20) -- (rel axis cs:-0.37,0.20) --
    (rel axis cs:-0.37,0.31) -- (rel axis cs:-0.34,0.31);

\draw[mpgreen, thick, xshift=-1.7cm]
    (rel axis cs:-0.34,0.01) -- (rel axis cs:-0.37,0.01) --
    (rel axis cs:-0.37,0.18) -- (rel axis cs:-0.34,0.18);

\end{axis}
\end{tikzpicture}%
}\\[2pt]
{\small (b) Task category distribution}
\end{minipage}%
\hfill
% ============================================================
% (c) Capability tag coverage (multi-label)
%   - changed to bars
%   - shorter horizontal plot width
% ============================================================
\begin{minipage}[t]{0.40\linewidth}
\centering
\resizebox{\linewidth}{!}{%
\begin{tikzpicture}
\begin{axis}[
    width=2cm,
    height=3.5cm,
    scale only axis,
    xbar,
    xmin=0,
    xmax=80,
    ymin=-0.6,
    ymax=11.6,
    enlarge y limits=0.05,
    axis x line*=bottom,
    axis y line*=left,
    y axis line style={draw=none},
    xlabel={\# of tasks with tag},
    xlabel style={font=\tiny},
    xtick={0,20,40,60,80},
    xticklabel style={font=\tiny},
    ytick={-0.3,0.6,1.6,2.6,3.6,5.4,6.4,7.4,8.4,9.4,10.4,11.4},
    yticklabel pos=left,
    yticklabels={
        Reference Resolution,
        Spatial Reasoning,
        Cross-File Comparison,
        Temporal Localization,
        Audio-Visual Alignment,
        Music Understanding,
        Speech Prosody,
        Speaker/Voice Identity,
        Non-Speech Audio,
        On-Screen Text,
        Speech Understanding,
        Visual Perception
    },
    yticklabel style={
        font=\tiny,
        anchor=east,
        align=right,
        xshift=1pt
    },
    ytick style={draw=none},
    xtick style={draw=none},
    xmajorgrids=true,
    ymajorgrids=false,
    grid style={dashed, gray!30},
    bar width=4.1pt,
    nodes near coords,
    every node near coord/.append style={
        font=\tiny,
        scale=0.8,
        anchor=west,
        xshift=1pt
    },
    clip=false
]

% Reasoning bars
\addplot[draw=none, fill=mpred] coordinates {
    (7,0)
    (12,1)
    (21,2)
    (44,3)
    (55,4)
};

% Perception bars
\addplot[draw=none, fill=mpblue] coordinates {
    (11,5)
    (13,6)
    (13,7)
    (26,8)
    (42,9)
    (57,10)
    (68,11)
};

% Group labels (left, vertical)
\node[
    rotate=90,
    anchor=center,
    font=\bfseries\tiny,
    text=mpblue,
    yshift=1.4cm,
    xshift=-0.05cm
] at (rel axis cs:-0.62,0.73) {Perception};

\node[
    rotate=90,
    anchor=center,
    font=\bfseries\tiny,
    text=mpred,
    yshift=1.4cm,
    xshift=0.1cm
] at (rel axis cs:-0.62,0.18) {Reasoning};

% Group brackets (between group labels and yticklabels)
\draw[mpblue, thick, xshift=-1.7cm]
    (rel axis cs:-0.34,0.46) -- (rel axis cs:-0.37,0.46) --
    (rel axis cs:-0.37,0.97) -- (rel axis cs:-0.34,0.97);

\draw[mpred, thick, xshift=-1.7cm]
    (rel axis cs:-0.34,0.01) -- (rel axis cs:-0.37,0.01) --
    (rel axis cs:-0.37,0.40) -- (rel axis cs:-0.34,0.40);

\end{axis}
\end{tikzpicture}%
}\\[2pt]
{\small (c) Capability tag distribution (multi-label)}
\end{minipage}
\hfill
\caption{
\textbf{Construction pipeline and statistics of \benchmark{}.}
(a) We curated 163 workflow-backed candidate scenarios and adapt them into Harbor tasks with license-compatible substitute multimedia files.
Successive automated validation, baseline review, and manual validation stages revise, refine, and prune the candidates, yielding a final suite of 105 tasks.
% (a) We start from 163 workflow-backed candidate tasks and implement them with license-compatible media. 
% Automated checks, baseline review, and manual validation filter and revise them into a final suite of 105 tasks.
(b) \benchmark{} encompasses 5 meta-categories and 16 fine-grained categories, representing industrial, academic, and research workflows.
(c) Distribution of Multi-label capability tags across tasks. 
%
\iffalse
\textbf{Construction pipeline and suite coverage of \benchmark{}.}
(a) Candidate tasks begin as workflow-backed ideas, are scoped, scaffolded in
Harbor, instantiated with license-compatible or controlled-synthesis media
assets, and filtered through automated validation and baseline/review screens,
yielding 105 released tasks from 163 consolidated candidates. (b) Workflow
taxonomy of the released suite: 16 fine-grained categories are grouped under
five workflow domains. (c) Multi-label capability coverage across tasks. Counts
and percentages in (c) are marginal tag frequencies, so they do not sum to 105
or 100\%.
\fi
}
\label{fig:mmtb-construction-coverage}
\vspace{-3mm}
\end{figure*}

%% file: 04evaluation_setup.tex
\section{Evaluation Setup: Harnesses, Models, Protocol, and Metrics}
\label{sec:evaluation-setup}
In this section, we describe the evaluation setup for \benchmark{}. We first introduce the harnesses and corresponding models used in our evaluation, including controlled Terminus-family variants and off-the-shelf terminal-agent baselines. We then present the shared execution protocol used across runs. Across these configurations, we compare agent performance using success and cost metrics.
For reproducibility, we provide the harness code and evaluation code online
\footnote{Code repository: \url{https://github.com/mm-tbench/multimedia-terminal-bench}.}.

\subsection{Harnesses and Models}
\label{sec:harnesses}

\begin{wraptable}{r}{0.44\linewidth}
\vspace{-2.50em}
\centering
\small
\setlength{\tabcolsep}{2pt}
\caption{
\textbf{Controlled Terminus harnesses.} 
Text, Image, Audio, and Video denote harness-level native perception tools.
% \emph{Terminal} denotes the shared shell command tool; image, audio, and video denote harness-level native perception tools.
}
\label{tab:harnesses}
\centering\resizebox{\linewidth}{!}{
\begin{tabular}{lcccc}
\toprule
Harness & Text & Image & Audio & Video \\
\midrule
Terminus-2~\cite{merrill2026terminalbench}     & \cmark & \xmark  & \xmark  & \xmark  \\
Terminus-KIRA~\cite{kira2024}  & \cmark & \cmark & \xmark  & \xmark  \\
% Terminus-A    & \cmark & \xmark  & \cmark  & \xmark  \\
% Terminus-V    & \cmark & \xmark  & \xmark  & \cmark  \\
Terminus-IA   & \cmark & \cmark  & \cmark  & \xmark  \\
Terminus-IV   & \cmark & \cmark  & \xmark  & \cmark  \\
Terminus-MM    & \cmark & \cmark  & \cmark  & \cmark  \\
\bottomrule
\end{tabular}
}
\vspace{-3.0em}
\end{wraptable}

\paragraph{Terminus family.}
Terminus-2~\citep{merrill2026terminalbench} is a minimal harness in which the agent interacts with a terminal, issues Bash commands, observes command outputs, and manipulates files through the filesystem. This terminal-only design is extended by Terminus-KIRA~\citep{kira2024}, which adds native image access and allows the agent to directly inspect images. Building on these two harnesses, we construct a family of controlled Terminus variants for \benchmark{}.
% or frames extracted from video.

The resulting variants are shown in Table~\ref{tab:harnesses}. All variants share the same terminal loop, task interface, and filesystem-based workflow, but differ in the subset of harness-level native perception tools exposed to the agent. We define a native perception tool as a harness interface that enables direct inspection of image, audio, or video content, rather than requiring the agent to first convert the content into an intermediate representation through shell commands. This design supports controlled ablations over native image, audio, and video access. Among these variants, Terminus-MM provides the full multimedia setting by exposing native perception tools for all three modalities. Terminus-MM also applies modality masking before model inference for each task. The harness scans the initial workspace, maps file extensions to available media modalities, and exposes only perception tools supported by the files present in that workspace.
For the controlled Terminus variants, we use four model backbones. Qwen3.5-122B~\citep{qwen3.5} and GPT-5.2~\citep{openai2026gpt52} cover text and image settings, while Gemini-2.5-Flash~\citep{comanici2025gemini} and Gemini-3.1-Pro~\citep{google2026gemini31pro} support the audio and video ablations.

% For these controlled Terminus variants, we use four model backbones: Qwen3.5-122B and GPT-5.2 as text-image references, and Gemini-2.5-Flash and Gemini-3.1-Pro as multimedia references for audio and video ablations. Implementation details are provided in Appendix~\ref{app:harness-details}, and full model identifiers, endpoint versions, inference settings, and pricing details are provided in Appendix~\ref{sec:appendix-models}.

\paragraph{Off-the-shelf terminal agents.}
Beyond the controlled Terminus family, \benchmark{} is also evaluated with Codex CLI~\citep{openai2026codexcli} and Claude Code~\citep{anthropic2026claudecode}, two off-the-shelf agents for command-line workflows. These systems operate through their own agent loops, tool interfaces, and media-handling mechanisms, rather than through the controlled Terminus harnesses. For evaluation, we instantiate Codex CLI with GPT-5.2 and Claude Code with Sonnet-4.6~\citep{anthropic2026sonnet46}. Together, these baselines provide a practical comparison point for assessing how existing terminal agents handle multimedia-file tasks under the same benchmark protocol. Details about harness implementations are provided in Appendix~\ref{app:harness-details}, with model and inference settings, endpoint versions, and pricing in Appendix~\ref{sec:appendix-models}.

\subsection{Execution Protocol}
\label{sec:execution-protocol}

To compare harness and model configurations under a common setting, all agents are evaluated in the same Harbor-style task environment. Each task provides an instruction, a persistent filesystem containing multimedia files, and one or more task-specified output paths. The agent operates within this workspace and must write the required output file or files to the specified paths. Depending on the task, the expected output may be a text file containing a selected filename, timestamp, interval, JSON record, CSV table, or edit list, or a generated multimedia artifact such as an edited clip. The evaluator reads the submitted output files and scores their contents and derived metadata.

Across all agents, the benchmark fixes the task workspace, preinstalled terminal tools, task instructions, evaluators, logging protocol, and 10-minute interaction budget. Agents may use ordinary terminal tools such as \texttt{ffmpeg}, \texttt{ffprobe}, speech transcription, OCR, silence detection, and generated signal-processing scripts to inspect multimedia files indirectly when native perception is unavailable. We allow these command-line workflows since they reflect realistic multimedia-workflow behavior. At the same time, we log commands, perception calls, intermediate files, and final artifacts, allowing the analysis to distinguish runs that use native media access from runs that inspect multimedia files through command-line workflows.

\subsection{Metrics}
\label{sec:metrics}

\paragraph{Success metrics.}
Let $T_i$ denote the $i$-th task among $N$ tasks, and let $A=(H,M)$ denote an agent configuration consisting of harness $H$ and language model $M$. Running agent $A$ on $T_i$ transforms the initial workspace into a final workspace state $y_i$, which includes the generated outputs. A task-specific verifier $V_i$ then evaluates this final state and assigns a partial score $s_i = V_i(y_i; A, T_i)$, where $s_i \in [0,1]$. Here, evaluation depends only on the final state $y_i$, not on the agent's intermediate actions, reasoning, or trajectory. Across the task set $\mathcal{T}=\{T_i\}_{i=1}^{N}$, we compute the binary success rate and partial success rate as follows:
\begin{equation}
\textsc{Binary}(A;\mathcal{T}) =
\frac{1}{N}\sum_{i=1}^{N} \mathbb{I}\left[s_i \geq \tau_i\right],
\qquad
\textsc{Partial}(A;\mathcal{T}) =
\frac{1}{N}\sum_{i=1}^{N} s_i,
\end{equation}
where $\tau_i$ is the task-specific acceptance threshold.
Together, \textsc{Binary} and \textsc{Partial} evaluate agent configuration $A$ by measuring the fraction of tasks that pass the verifier threshold and the average partial correctness, respectively.

\paragraph{Cost metrics.}
In addition to success metrics, we report mean API cost per task and mean agent execution time per task. These metrics capture the practical efficiency of each agent configuration, since terminal agents must not only produce correct outputs but also do so within reasonable cost and time. Details of the cost computation and time boundaries are provided in Appendix~\ref{app:cost-methodology}.

%% file: 05main_results.tex
\section{Results and Analyses}
\label{sec:results}

\input{tex-table-figure/modality_table}

\input{tex-table-figure/proxy_recon_table}

\input{tex-table-figure/harness_routing_table}
\input{tex-table-figure/mm_codex_overlap_figure}

\subsection{Native Multimedia Access Improves Multimedia-File Task Solving}

Table~\ref{tab:main-results} summarizes the main \benchmark results.
The results indicate that text-only and text-image access are insufficient for
many multimedia-file tasks.
On Gemini-3.1-Pro, text-only Terminus-2 reaches $0.124$ binary and $0.162$
partial success, while image-augmented Terminus-KIRA reaches $0.105$ binary
and $0.159$ partial success.
In contrast, adding native media access leads to substantially higher performance.
Terminus-IA and Terminus-IV each reach $0.333$ binary success, and Terminus-MM reaches $0.371$ binary and $0.469$ partial success.
Gemini-2.5-Flash shows the same qualitative ordering, with lower absolute
performance.
Beyond success rates, Terminus-MM is also cost-efficient: it has the lowest mean
API cost among Gemini-2.5-Flash agents and the second-lowest mean API cost
among Gemini-3.1-Pro agents.
We discuss this cost advantage further in Section~\ref{sec:result-cost}.

Table~\ref{tab:ablation-modality} further supports this pattern.
Audio-only and video-only native access already improve over text-only access;
adding image access on top of either provides additional gains; and the full
T+I+A+V setting obtains the best overall result.
Thus, image perception is useful as a complement to audio or video evidence, but
the main bottleneck in \benchmark is access to media cues such as speech, sound
events, motion, timing boundaries, and audio-visual alignment.
In other words, native multimedia-file access is an essential component of effective
task solving in \benchmark.

\subsection{{Command-Line Conversions Are Less Efficient Than} Native Access}
\label{sec:result-cost}

When native perception for a required modality is unavailable, agents inspect media through command-line conversion tools. Table~\ref{tab:proxy-cost} quantifies this overhead on matched co-success cases: tasks solved by both partial-modality harnesses and Terminus-MM, where the partial harness lacks a required modality and attempts to recover it through terminal tools. We provide the filtering details in Appendix~\ref{app:proxy-cost}. We report overhead as ratios of USD API cost and trajectory turns relative to Terminus-MM.

Conversion-heavy successful runs incur substantially higher overhead: average API-cost ratios range from 1.63$\times$ to 7.72$\times$, with worst cases reaching 30.11$\times$ when native video is missing and 42.49$\times$ when native audio is missing. Turn ratios also increase in most settings, with a worst case of 8.10$\times$. Because failed and timed-out conversion attempts are excluded, these ratios characterize the cost of successful conversions.

This overhead stems from a longer evidence-acquisition path: the agent must choose an intermediate representation, run the corresponding tools, interpret lossy derived evidence, and often retry before producing the final artifact. Native access shortens this path by letting the agent inspect raw media directly, while terminal commands remain necessary for artifact construction.

\subsection{Multimedia Terminal Harnesses Need Both Native Access and Tool-use Ability}
\label{sec:results-harness}
Figure~\ref{fig:mm-codex-venn} shows that Terminus-MM and Codex CLI solve overlapping but non-nested task sets across the 105-task suite: 28 tasks are solved only by Terminus-MM, 6 only by Codex CLI, 11 by both, and 60 by neither. 
The Terminus-MM-only and Codex-CLI-only task sets suggest different strengths of the two systems. Terminus-MM-only tasks tend to require native modality understanding: the agent listens to audio or watches video directly and grounds its answer in that perceptual evidence. Codex-CLI-only tasks tend to be cases where command-line conversion tools are sufficient: the tools turn media into text or numeric evidence that the agent can reason over. A dedicated harness for multimedia-file tasks therefore needs both native access and tool-use ability. Appendix~\ref{app:regime-analysis} enumerates the detailed partition.

However, native access alone is not enough: the harness must also decide which perception tools to expose. Table~\ref{tab:tool-exposure} compares Terminus-MM against Terminus-MM w/o modality masking. Removing the mask lowers both binary and partial success on each backbone, suggesting that an unmasked tool list can draw the agent into redundant evidence gathering over unnecessary modalities. Trajectory excerpts are provided in Appendix~\ref{app:tool-exposure}. This observation suggests that modality selection is the responsibility of a dedicated multimedia terminal harness.

\subsection{Failure Modes Differ between Native-Media and Terminal Agents}
\label{sec:result-failure}

We label every binary-failed run for the two agents: 66 failures from Terminus-MM and 88 failures from Codex CLI.
Failure analysis shows that Terminus-MM and Codex CLI fail in different parts of the workflow. 
We label each binary-failed run with primary failure signature to identify the stage and failure points.
\textit{Timeout (tool setup)} denotes runs that exhaust the budget while preparing the additional tool environment. 
\textit{Timeout (tool execution)} denotes runs that timeout while running or retrying the actual multimedia-processing tool. 
\textit{Wrong (output format)} denotes runs that produce an artifact with the wrong file format or the artifact is submitted in different way. 
\textit{Wrong (wrong approach)} denotes runs whose overall plan is incompatible with the task goal. 
\textit{Wrong (correct approach, low precision)} denotes runs that identify the correct type of evidence and use a plausible approach, but the tool results in a temporal interval, label, or threshold that falls outside the verifier tolerance.
\textit{Wrong (tool failure)} denotes runs where an appropriate tool is invoked but crashes, returns unusable output, or produces an error that the agent fails to recover from. 
\textit{Wrong (model reasoning)} denotes runs where the relevant evidence is available to the agent, either through native perception or tool outputs, but the model maps that evidence to the wrong decision or artifact content.

Figure~\ref{fig:failure-signatures} shows that reliance on multimedia conversion tools shifts failures toward tool-mediated workflow errors. Terminus-MM still has a large model-reasoning component: 47\% of its failed runs end with the model misinterpreting or misusing available evidence. However, tool-operation failures remain relatively smaller for Terminus-MM: tool setup, tool execution, and tool failure together account for 24\%. By contrast, Codex CLI shows a much larger tool-use failure footprint. Setup and execution timeouts alone account for 39\% of its failures, and adding tool failures and low precision raises the strict tool-operation share to about 47\%. 

This observation suggests that missing native audio-video perception forces the agent to externalize perceptual evidences through a longer terminal pipeline. 
An agent without native multimedia access must choose which intermediate representation to build, invoke the corresponding tool, wait for the tool to finish, recover from tool errors, interpret a lossy results, and finally commit the result in the exact artifact format expected by the verifier. Each additional step creates another point of failure before the model can make the final media-grounded decision and commit it as a verifier-compatible artifact.
Fully closing this gap requires agents that can combine the efficiency gains of native multimedia access, the precision and controllability gains of reliable terminal-tool use, and stronger model-level reasoning for selecting evidence and translating it into well-formatted artifacts.

%% file: tex-table-figure/modality_table.tex
\begin{table*}[t]
\centering
\small
\caption{
\textbf{\benchmark{} results across different harnesses and model backbones.}
Within a given backbone, full modality access yields substantial gains in success rates.
\textbf{Bold} and \underline{underline} denote the best and second-best results for each backbone, respectively. T, I, A, and V denote text, image, audio, and video access.
%
\iffalse
\textbf{Main results across backbones.}
Access codes: T = terminal,
I = image, A = audio, V = video. \textsc{Terminus-MM} uses a workspace-routed
schema (\S\ref{sec:result-tool-exposure}); other rows use a static schema.
Bold and underline mark the best and second-best in each success-rate column.
Single-modality (A, V) and AV variants are in
Table~\ref{tab:ablation-modality}.
Time = mean agent execution wall (inside \texttt{agent.run()}; includes LLM
API latency, tool-call execution, and perception-tool latency; excludes
container setup and verifier scoring).
\fi
}
\label{tab:main-results}
\begin{tabular}{lllcc|cc}
\toprule
\multirow{2}{*}{Harness} &
\multirow{2}{*}{Model} &
\multirow{2}{*}{\makecell{Modality \\ Access}} &
\multicolumn{2}{c|}{Success Rate $\uparrow$} &
\multicolumn{2}{c}{Cost $\downarrow$} \\
& & & Binary & Partial & API & Time \\
\midrule
Terminus-2~\cite{merrill2026terminalbench}    & \multirow{2}{*}{Qwen3.5-122B} & T   & 0.105 & 0.159 & \$0.101 & 510s \\
Terminus-KIRA~\cite{kira2024} &                                 & T+I & 0.095 & 0.165 & \$0.233 & 519s \\
\midrule
Terminus-2~\cite{merrill2026terminalbench}    & \multirow{2}{*}{GPT-5.2}       & T   & 0.105 & 0.149 & \$0.818 & 500s \\
Terminus-KIRA~\cite{kira2024} &                                 & T+I & 0.114 & 0.150 & \$1.672 & 540s \\
\midrule
Terminus-2~\cite{merrill2026terminalbench}     & \multirow{5}{*}{Gemini-2.5-Flash} & T        & 0.067 & 0.136 & \underline{\$0.115} & \underline{248s} \\
Terminus-KIRA~\cite{kira2024}  &                                    & T+I      & 0.067 & 0.129 & \$0.226 & 290s \\
Terminus-IA   &                                    & T+I+A    & 0.133 & 0.181 & \$0.234 & 269s \\
Terminus-IV   &                                    & T+I+V    & \underline{0.162} & \underline{0.222} & \$0.184 & 272s \\
\rowcolor{lightgray}
Terminus-MM    &                                    & T+I+A+V  & \textbf{0.229} & \textbf{0.305} & \textbf{\$0.099} & \textbf{229s} \\
\midrule
Terminus-2~\cite{merrill2026terminalbench}     & \multirow{5}{*}{Gemini-3.1-Pro} & T        & 0.124 & 0.162 & \textbf{\$0.772} & 538s \\
Terminus-KIRA~\cite{kira2024}  &                                  & T+I      & 0.105 & 0.159 & \$2.061 & 544s \\
Terminus-IA   &                                  & T+I+A    & \underline{0.333} & 0.406 & \$1.742 & 460s \\
Terminus-IV   &                                  & T+I+V    & \underline{0.333} & \underline{0.432} & \$1.283 & \textbf{434s} \\
\rowcolor{lightgray}
Terminus-MM    &                                  & T+I+A+V  & \textbf{0.371} & \textbf{0.469} & \underline{\$1.228} & \underline{442s} \\
\midrule
Claude Code & Sonnet-4.6 & T+I & 0.162 & 0.186 & \$1.735 & 516s \\
Codex CLI   & GPT-5.2    & T+I & 0.162 & 0.202 & \$7.117 & 529s \\
\bottomrule
\end{tabular}
\end{table*}

%%%%
\begin{table*}[t]
\centering
\small
\caption{
\textbf{Modality-ladder ablation on Gemini-3.1-Pro.} 
single non-text modality (A, V) $\to$ image-augmented (IA, IV) $\to$ full \textsc{MM}.
% Convention as in Table~\ref{tab:main-results}.
% Time = mean agent execution wall (inside \texttt{agent.run()}; includes LLM
% API latency, tool-call execution, and perception-tool latency; excludes
% container setup and verifier scoring).
}
\label{tab:ablation-modality}
\begin{tabular}{llcc|cc}
\toprule
\multirow{2}{*}{Harness} &
\multirow{2}{*}{\makecell{Modality \\ Access}} &
\multicolumn{2}{c|}{Success Rate $\uparrow$} &
\multicolumn{2}{c}{Cost $\downarrow$} \\
& & Binary & Partial & API & Time \\
\midrule
Terminus-A    & T+A      & 0.257 & 0.349 & \$1.992 & 493s \\
Terminus-V    & T+V      & 0.286 & 0.395 & \$1.118 & 417s \\
\midrule
Terminus-IA   & T+I+A    & 0.333 & 0.406 & \$1.742 & 460s \\
Terminus-IV   & T+I+V    & 0.333 & 0.432 & \$1.283 & 434s \\
\midrule
\rowcolor{lightgray}
Terminus-MM   & T+I+A+V  & \textbf{0.371} & \textbf{0.469} & \$1.228 & 442s \\
\bottomrule
\end{tabular}
\end{table*}

%% file: tex-table-figure/proxy_recon_table.tex
\begin{table*}[t]
\centering
\small
\setlength{\tabcolsep}{12pt}
\renewcommand{\arraystretch}{1.12}
\caption{
\textbf{{Overhead of inspecting multimedia files through command-line conversion tools.}}
We compare partial-modality harnesses with {Terminus-MM} using the Gemini-3.1-Pro backbone. 
To isolate the overhead of inspecting multimedia files through command-line tools rather than using native modality access,
% To isolate the overhead \textcolor{blue}{an agent incurs when it inspects media through command-line tools }, 
we filter for tasks where both harnesses succeed and the required modality is inaccessible to the partial-modality harness.
$n$ denotes the number of filtered tasks.
Ratios represent the cost and turn counts of the evaluated harness divided by those of {Terminus-MM}.
%
\iffalse
\textbf{proxy-reconstruction overhead in matched comparisons against \textsc{Terminus-MM}
on the same Gemini-3.1-Pro backbone.} Ratios are partial-access harness divided
by \textsc{Terminus-MM}, computed only on tasks both harnesses solve. Seven
Terminus rows cover every same-backbone partial-access variant with a
non-empty matched set.
$n$ is the final matched set after filtering for co-success, missing-modality
requirement, and observed proxy reconstruction. Avg.\ compares mean API cost
or mean trajectory turns over that matched set; Worst reports the largest
per-task ratio.
\fi
}
\label{tab:proxy-cost}
\resizebox{0.95\textwidth}{!}{%
\begin{tabular*}{\textwidth}{llccccc}
\toprule
\multirow[c]{2}{*}{Harness}
& \multirow[c]{2}{*}{Modality Access}
& \multirow[c]{2}{*}{$n$}
& \multicolumn{2}{c}{API cost ratio}
& \multicolumn{2}{c}{Turn ratio} \\
\cmidrule(lr){4-5}\cmidrule(lr){6-7}
& & & Avg. & Worst & Avg. & Worst \\
\midrule
Terminus-2
    & T
    & 7
    & 4.12$\times$
    & 26.48$\times$
    & 1.00$\times$
    & 2.71$\times$ \\
Terminus-KIRA
    & T+I
    & 4
    & 1.63$\times$
    & 3.40$\times$
    & 1.39$\times$
    & 2.00$\times$ \\
Terminus-A
    & T+A
    & 14
    & 4.38$\times$
    & 19.37$\times$
    & 1.77$\times$
    & 3.83$\times$ \\
Terminus-V
    & T+V
    & 13
    & 1.84$\times$
    & 11.64$\times$
    & 1.15$\times$
    & 2.17$\times$ \\
Terminus-IA
    & T+I+A
    & 13
    & 4.20$\times$
    & 30.11$\times$
    & 2.21$\times$
    & 8.10$\times$ \\
Terminus-IV
    & T+I+V
    & 14
    & 7.72$\times$
    & 42.49$\times$
    & 2.01$\times$
    & 5.88$\times$ \\
Terminus-AV
    & T+A+V
    & 4
    & 2.00$\times$
    & 6.08$\times$
    & 1.28$\times$
    & 2.06$\times$ \\
\bottomrule
\end{tabular*}
}

% \vspace{0.25em}
\end{table*}

%% file: tex-table-figure/harness_routing_table.tex
% \begin{table*}[t]
% \centering
% \small
% \caption{
% \textbf{Ablation study over modality masking.}
% }
% \label{tab:tool-exposure}
% \centering\resizebox{\linewidth}{!}{
% \begin{tabular}{@{}llcc@{}}
% \toprule
% Backbone & Harness & Binary Success Rate $\uparrow$ & Partial Success Rate $\uparrow$ \\
% \midrule
% Gemini-2.5-Flash
%     & {Terminus-MM w/o modality masking} & 0.171 & 0.267 \\
%     & {Terminus-MM}   & \textbf{0.229} & \textbf{0.305} \\
% \midrule
% Gemini-3.1-Pro
%     & {Terminus-MM w/o modality masking} & 0.324 & 0.426 \\
%     & {Terminus-MM}   & \textbf{0.371} & \textbf{0.469} \\
% \bottomrule
% \end{tabular}
% }
% \end{table*}

\begin{table*}[t]
\centering
\small
\caption{
\textbf{Ablation study over modality masking.}
}
\label{tab:tool-exposure}
\resizebox{0.95\linewidth}{!}{
\begin{tabular}{llcc}
\toprule
Backbone & { Harness} & Binary Success Rate $\uparrow$ & Partial Success Rate $\uparrow$ \\
\midrule
\multirow{2}{*}{Gemini-2.5-Flash}
    & { Terminus-MM w/o modality masking} & 0.171 & 0.267 \\
    & \cellcolor{lightgray}
{Terminus-MM}   & \cellcolor{lightgray}
\textbf{0.229} & \cellcolor{lightgray}
\textbf{0.305} \\
\midrule
\multirow{2}{*}{Gemini-3.1-Pro}
    & { Terminus-MM w/o modality masking} & 0.324 & 0.426 \\
    & \cellcolor{lightgray}
{Terminus-MM}   & \cellcolor{lightgray}
\textbf{0.371} & \cellcolor{lightgray}
\textbf{0.469} \\
\bottomrule
\end{tabular}
}
\end{table*}

%% file: tex-table-figure/mm_codex_overlap_figure.tex
\begin{figure*}[t]
\centering
\begin{minipage}[t]{0.48\linewidth} 
\centering
\includegraphics[width=0.9\linewidth]{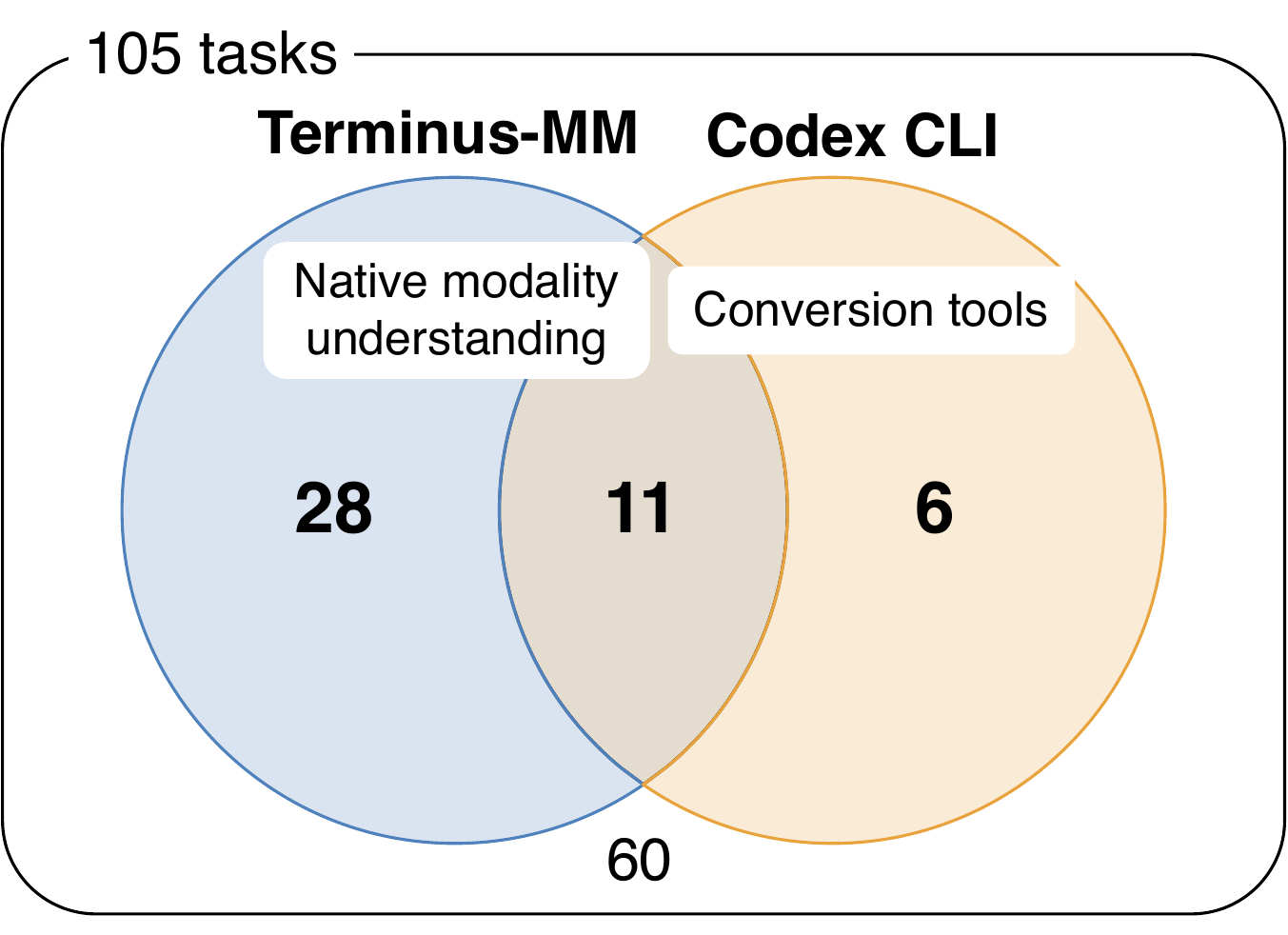}
\caption{
\textbf{Overlap of solved tasks across Terminus-MM and Codex CLI.}
The non-overlapping regions indicate task subsets for which different capabilities are useful for successful task completion.
% Solved tasks indicate tasks solvable by native modality understanding and conversion tools.
}
\label{fig:mm-codex-venn}
\end{minipage}
\hfill
\begin{minipage}[t]{0.48\linewidth} 
\input{tex-table-figure/failure_analysis_figure}
\end{minipage}
\vspace{-3mm}
\end{figure*}

%% file: tex-table-figure/failure_analysis_figure.tex
% \begin{figure}[t]
  \includegraphics[width=\linewidth]{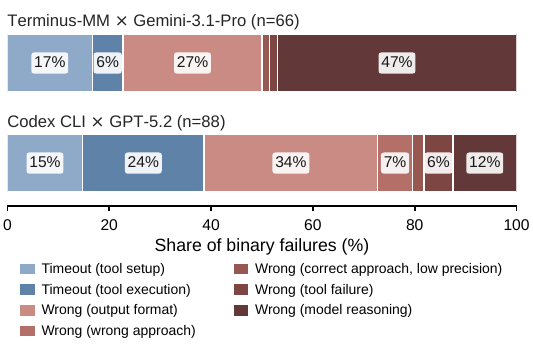}
  \caption{
    \textbf{Failure signatures for failed runs of {Terminus-MM} and Codex CLI. }
    Percentages are normalized over failed tasks for each agent.
  }
  \label{fig:failure-signatures}
% \end{figure}

%% file: 07related_work.tex
\section{Related Work}
\label{sec:related}

\paragraph{Multimodal perception benchmarks.}
MMMU, Video-MME, JointAVBench, Video-Holmes, and
AVTrustBench~\citep{yue2024mmmu,fu2024videomme,chao2025jointavbench,
cheng2025videoholmes,chowdhury2025avtrustbench} reduce media
understanding to a text answer. They evaluate \emph{what the model
sees}, not what it does with stored media in a workflow.

\paragraph{Computer-use and terminal-agent benchmarks.}
WebArena, VisualWebArena, OSWorld, and
Terminal-Bench~\citep{zhou2024webarena,koh2024visualwebarena,
xie2024osworld,merrill2026terminalbench} evaluate agency and
artifact production but rarely make media content the object of
work. Adjacent settings such as VideoWebArena and
OmniPlay~\citep{jang2024videowebarena,bie2025omniplay} couple long
video context to web or game agents, but target embodied or
interactive surfaces rather than persistent filesystem workflows.

\paragraph{Artifact-level evaluation and benchmark hygiene.}
SWE-bench, MLE-bench, and
AppWorld~\citep{jimenez2024swebench,chan2024mlebench,
trivedi2024appworld} established artifact-level scoring but on
code, models, and app state rather than stored media. \benchmark{} sits at the intersection of these three lines: agents
work on stored multimedia files in a terminal and are scored on the
final workspace state they produce. The shortcut-as-data
framing extends prior work on benchmark hygiene -- VQA shortcut
analyses~\citep{goyal2017vqashortcut} and vision-language
contamination studies~\citep{chen2024mmstar} -- by treating
{an agent's command-line workarounds for a missing modality} as evidence
{about the workflow rather than} contamination to suppress.

%% file: 08discussion.tex
\section{Discussion}
\label{sec:discussion}

\paragraph{Human baselines are not directly comparable.}
% \benchmark{} measures agent performance against artifact-level verifiers,
% with no human reference. A like-for-like baseline is itself non-trivial:
% human experts rarely solve multimedia-file tasks in a terminal, and
% instead rely on professional GUI applications (Premiere, Audacity,
% DaVinci Resolve, Photoshop) with interactive timelines, scrubbing, and
% layer panels. Restricting experts to the terminal yields an artificially
% low ceiling; allowing their native tools compares two different
% interaction surfaces. Without such a study we cannot bound how close
% agents are to human performance, and the 60 tasks in the shared-failure
% set of Section\ref{sec:result-harness} cannot be classified as ``hard for
% agents but easy for humans''. We leave a careful human-baseline design
% to future work.
\benchmark{} measures agent performance with artifact-level verifiers, without a human reference. A like-for-like baseline is non-trivial: human experts rarely solve multimedia-file tasks in a terminal, instead relying on professional GUI tools such as Premiere, Audacity, DaVinci Resolve, or Photoshop, with interactive timelines, scrubbing, and layer panels. Restricting experts to the terminal would impose an artificial ceiling, while allowing native tools would compare different interaction surfaces. Thus, without a dedicated study, we cannot determine how close agents are to human performance or whether the 60 shared-failure tasks in Section~\ref{sec:results-harness} are hard for agents but easy for humans. We leave careful human-baseline design to future work.

\paragraph{Results on an extended budget remains unmeasured.}
All evaluated agents share a 600-second per-task wall-clock budget, and
we do not sweep it. The over-checking and tool/setup-loop failures in
Section~\ref{sec:result-failure} may close at longer budgets, especially for
Codex CLI and Claude Code, which show high tail-token consumption near
the cap; the wrong-evidence and lossy-analysis classes likely persist
regardless. A budget sweep would let future work separate these regimes.

\paragraph{Broader impact.}
% \benchmark{} is an evaluation benchmark, not a deployable system: its
% released artifacts are tasks, instructions, and verifiers built from
% license-cleared and synthetic media. Stronger terminal agents on
% \benchmark{} can be used in beneficial workflows (accessibility QA,
% content production, compliance review) and in harmful ones (automated
% surveillance, large-scale media tampering); the benchmark itself does
% not introduce new generative or extraction capabilities. Practitioners
% deploying multimedia terminal agents should pair capability claims with
% explicit license, consent, and audit-trail policies.
% \benchmark{} is an evaluation benchmark rather than a deployable system: its released artifacts consist of tasks, instructions, and verifiers built from license-cleared and synthetic media. Stronger terminal agents on \benchmark{} may support beneficial workflows, such as accessibility QA, content production, and compliance review, but may also enable harmful ones, such as automated surveillance or large-scale media tampering. The benchmark itself introduces no new generative or extraction capabilities. Deployments of multimedia terminal agents should therefore pair capability claims with explicit policies for license, consent, and audit trails.

As a benchmark for multimedia-file workflows in persistent terminal workspaces, MMTB supports progress toward agents that can inspect, transform, validate, and produce outputs from multimedia files. 
Progress on MMTB can benefit the growing number of users who rely on terminal agents such as Claude Code and Codex CLI to automate file-based workflows involving multimedia assets, including media production, audio-video analysis, quality control, and structured annotation. 
MMTB also provides a shared evaluation ground for researchers developing omni-modal models and multimedia terminal-agent harnesses. 
This opens a fairer venue for comparing systems under common tasks and evaluators, and for attributing improvements to native multimedia access, reliable tool use, artifact construction, or model-level reasoning.

%% file: 09appendix.tex
\section{Implementation Details}

\subsection{Harness Implementation Details}
\label{app:harness-details}

This appendix supports Section~\ref{sec:harnesses}. Table~\ref{tab:harness-details}
summarizes the seven Terminus-family harnesses evaluated in the paper, listing
their native perception tools, tool-routing policy, prompt template, and class
definition. Algorithm~\ref{alg:v5-routing} states the workspace-aware tool
routing used by Terminus-MM; the remaining harnesses use a static schema.

\begin{table*}[h]
\centering
\small
\caption{Harness implementations evaluated in this paper. Tool routing is
\emph{static} (the perception schema is fixed at construction) or
\emph{dynamic} (the schema is re-derived per task at run start from a
workspace scan).}
\label{tab:harness-details}
\begin{tabular}{@{}lll@{}}
\toprule
Harness & Perception tools & Routing \\
\midrule
Terminus-2     & ---                                          & static  \\
Terminus-KIRA  & view\_image                                  & static  \\
Terminus-A     & listen\_audio                                & static  \\
Terminus-IA    & view\_image, listen\_audio                   & static  \\
Terminus-IV    & view\_image, watch\_video                    & static  \\
Terminus-MM w/o modality masking & view\_image, listen\_audio, watch\_video & static  \\
Terminus-MM    & subset of \{view\_image, listen\_audio, watch\_video\} & dynamic \\
\bottomrule
\end{tabular}
\end{table*}

\paragraph{Workspace-aware tool routing (Terminus-MM).}
The \emph{static} harnesses fix their perception schema at construction time.
Terminus-MM instead derives the schema once per task at run start by scanning
the workspace, mapping file extensions to perception modalities, and keeping
only the perception tools whose target modality is present in the workspace
(Algorithm~\ref{alg:v5-routing}). The \texttt{view\_image} keep rule is
``include \texttt{view\_image} whenever any media file is present'', even on
audio-only or video-only workspaces, so that frames or spectrograms produced
by \texttt{bash\_command} retain a visual perception path.

\begin{algorithm}[h]
\caption{Workspace-aware tool routing (Terminus-MM). Invoked once per task
at run start, before the first LLM call. The tool schema sent to the LLM is
the routed subset returned here.}
\label{alg:v5-routing}
\begin{algorithmic}[1]
\State files $\gets$ \textsc{ListFiles}(workspace\_dir, max\_depth=6)
\State modalities $\gets \emptyset$
\For{$f \in$ files}
    \State ext $\gets$ \textsc{Extension}($f$)
    \If{ext $\in$ \{.wav,.mp3,.ogg,.flac,.aac,.m4a\}} modalities $\gets$ modalities $\cup$ \{audio\} \EndIf
    \If{ext $\in$ \{.mp4,.webm,.avi,.mov,.mkv\}}      modalities $\gets$ modalities $\cup$ \{video\} \EndIf
    \If{ext $\in$ \{.png,.jpg,.jpeg,.gif,.webp\}}     modalities $\gets$ modalities $\cup$ \{image\} \EndIf
\EndFor
\State keep $\gets$ \{execute\_commands, task\_complete\}
\If{modalities $\neq \emptyset$}                  keep $\gets$ keep $\cup$ \{view\_image\}     \EndIf
\If{audio $\in$ modalities}                       keep $\gets$ keep $\cup$ \{listen\_audio\}   \EndIf
\If{video $\in$ modalities}                       keep $\gets$ keep $\cup$ \{watch\_video\}    \EndIf
\State \Return $\{t \in \mathrm{MM\_TOOLS} : t.\mathrm{name} \in \mathrm{keep}\}$
\end{algorithmic}
\end{algorithm}

\paragraph{Prompt-template structure.}
The MMTB prompt templates share a fixed skeleton: a system-role preamble, a
per-tool description block, an agent-constraints block (no human intervention;
minimal state changes before \texttt{task\_complete}), and \texttt{\{instruction\}}
plus \texttt{\{terminal\_state\}} placeholders that Harbor formats at task
start. The variants differ only in which per-tool description blocks are
present. The MM canonical lists all five tools (\texttt{execute\_commands},
\texttt{task\_complete}, \texttt{view\_image}, \texttt{watch\_video},
\texttt{listen\_audio}) with no disclaimer; it is used by Terminus-MM w/o
modality masking and is also the prompt deployed at runtime by Terminus-MM.
The Terminus-IA variant drops the \texttt{watch\_video} block and adds a
single ``You CANNOT call \texttt{watch\_video}'' line; Terminus-IV (drops
\texttt{listen\_audio}) and Terminus-A (drops both \texttt{view\_image} and
\texttt{watch\_video}) follow the same one-line disclaimer pattern.
Terminus-KIRA uses a structurally distinct Apache-2.0-vendored prompt that
predates the MM family.

The schema-routing component of Terminus-MM (Algorithm~\ref{alg:v5-routing})
is the load-bearing contribution analyzed in Section~\ref{sec:results-harness};
the deployed runtime prompt remains the MM canonical.

\subsection{Backbone Model Details}
\label{sec:appendix-models}

Model names in Table~\ref{tab:main-results} are abbreviations chosen for table density. Full identifiers,
provider routes, and version slugs for every model that appears in
the paper are in Table~\ref{tab:appendix-models}. The full per-cell
results table corresponding to the abbreviated main-text version is
reproduced as Table~\ref{tab:main-results-full}.

\begin{table*}[h]
\centering
\scriptsize
\setlength{\tabcolsep}{3pt}
\caption{Backbone model details. The ``In paper'' column gives the
abbreviation used in the main-text tables. The ``Full identifier''
column gives the canonical model name as listed by the provider.
The ``Route / version'' column gives the access path used for the
reported runs (provider slug routed via OpenRouter, native API, or
installable terminal agent). Runs queried on
\texttt{2026-04-29}--\texttt{2026-05-01} (UTC).}
\label{tab:appendix-models}
\begin{tabularx}{\linewidth}{L{3.0cm} L{4.6cm} Y c c}
\toprule
In paper & Full identifier & Route / version & Class & Type \\
\midrule
\multicolumn{5}{l}{\emph{Neutral-harness backbones}} \\
Qwen3.5-122B & Qwen3.5-122B-A10B & \texttt{qwen/qwen3.5-122b-a10b} via OpenRouter & VLM & open \\
Gemini-2.5-Flash & Google Gemini 2.5 Flash & \texttt{google/gemini-2.5-flash} via OpenRouter & Omni & closed \\
Gemini-3.1-Pro & Google Gemini 3.1 Pro Preview & \texttt{google/gemini-3.1-pro-preview} via OpenRouter & Omni & closed \\
\midrule
\multicolumn{5}{l}{\emph{Installable CLI-agent backbones (subscription-routed)}} \\
Sonnet-4.6 & Anthropic Claude Sonnet 4.6 & \texttt{anthropic/claude-sonnet-4-6} via Claude Code (Claude Max OAuth) & VLM & closed \\
GPT-5.2 & OpenAI GPT-5.2 & \texttt{openai/gpt-5.2} via Codex CLI (Codex Pro OAuth) & VLM & closed \\
\bottomrule
\end{tabularx}
\end{table*}

\input{tex-table-figure/modality_table_full}

\subsection{Cost and Time Measurement Methodology}
\label{app:cost-methodology}

This appendix specifies the API cost and time bases used in the §4 main
table and {in the cost ratios reported in
Table~\ref{tab:proxy-cost} (the cost of inspecting a missing modality through
command-line tools, relative to native perception on the same task).}

\paragraph{API cost.}
We report a uniform-proxy cost computed identically across all harnesses as
$\text{cost}_{\text{uniform}} = n_{\text{input}} \times r_{\text{input}}
+ n_{\text{output}} \times r_{\text{output}}$, where $r_{\text{input}}$ and
$r_{\text{output}}$ are the posted per-token list rates for the model and
$n_{\text{input}}$ is the total prompt tokens including any cached portion
(no prompt-cache discount applied). We chose this uniform basis because
prompt-cache token capture was asymmetric across harnesses in our sweep:
Terminus-2 (Harbor's built-in) recorded cached tokens via the OpenRouter
\texttt{usage.prompt\_tokens\_details.cached\_tokens} field, while the custom
Terminus subclasses read an Anthropic-style field name and therefore returned
zero on Gemini routes. Reporting cache-discounted billed cost would mix
real billed amounts (Terminus-2 only) with token-proxy estimates without
discount (other harnesses); the uniform-proxy basis treats every row as
un-cached at the same rate. The relative cross-harness ranking is preserved.
Absolute USD figures should be read as un-cached upper bounds; agents that
rely heavily on prompt caching at deployment (notably Codex CLI) will incur
proportionally less in production. Posted rates are recorded in the
released cost-aggregation script (snapshot 2026-04-29).

\paragraph{Time.}
We report the agent execution wall, measured by Harbor's \texttt{TimingInfo}
wrapper around the agent's run entry point in the Harbor trial runner.
This window includes LLM API latency,
tool-call execution, and perception-tool latency. It excludes container or
sandbox setup, harness initialization, and verifier
scoring. Time boundaries are identical across all evaluated harnesses
because they share the same \texttt{Trial} wrapper code path.

\subsection{Compute Resources}
\label{app:compute-resources}

Terminus harness sweeps are run on Daytona managed sandboxes with
inference routed through OpenRouter. Codex CLI and Claude Code are run
in a local Docker container with OAuth subscription authentication. All
inference is API-mediated; no local GPU is used.

The full paper-cited grid of 2{,}520 cells (24 (model, harness,
revision) triples $\times$ 105 tasks, single seed) takes approximately
30 wall-clock hours when sandboxes are dispatched in parallel. We
additionally ran preliminary or superseded experiments (multi-seed
pilots, replaced baselines, harness ablation variants) that do not
appear in the reported tables.

\section{Supplementary Analyses}

\subsection{Public Release, License, and Croissant Metadata}
\label{sec:appendix-release}

The benchmark is publicly hosted on Hugging Face as a single
dataset mirror linked from the supplementary material. Per-asset
media licenses are recorded in each task's \texttt{media.toml}
(predominantly CC-BY, CC0, and public-domain for source media; MIT
for benchmark code; see Section~\ref{sec:appendix-source-corpus} for the
full breakdown including a small number of CC-BY-NC and GPL files
covered by the license caveats below) and propagate into the
dataset's Croissant 1.0 metadata, which we hand-craft to include
all seven NeurIPS-required Responsible AI fields
(\texttt{rai:dataLimitations}, \texttt{rai:dataBiases},
\texttt{rai:personalSensitiveInformation},
\texttt{rai:dataUseCases}, \texttt{rai:dataSocialImpact},
\texttt{rai:hasSyntheticData}, \texttt{prov:wasGeneratedBy}) plus
Croissant-RAI extensions for collection, preprocessing,
annotation, and maintenance protocols. The Croissant file passes
all four checks of the NeurIPS Croissant validator (JSON format,
Croissant schema, record generation, RAI completeness). A
Gebru-style \texttt{DATASHEET.md} ships alongside the dataset and
covers motivation, composition, collection, preprocessing, uses,
distribution, and maintenance in the standard seven-section
narrative. Per-task self-contained subdirectories (1\,MB to
${\sim}300$\,MB of media each) serve as natural samples for
reviewers without requiring the full ${\sim}6$\,GB download.
% TODO(SUBMISSION): Replace with concrete URL pointers once the anonymised review channel is set up.

\subsection{Source Corpus, Language Coverage, Demographics}
\label{sec:appendix-source-corpus}

Source media is drawn predominantly from real recordings under
CC-BY, CC0, or public-domain licenses (NASA archives, MIT
OpenCourseWare, archive.org, Wikimedia Commons, Freesound).
Synthetic content is used only for targeted inserts where a
controlled distortion is required (e.g.\ short scripted
multi-speaker dialogues via a permissively licensed neural TTS);
synthesised material is never the dominant signal of a task.
The 105 tasks are predominantly English-speech; bilingual material
appears in localisation-flavoured tasks (German and French in
document subsets, dubbed content in subtitling tasks). Systematic
coverage of non-Latin scripts and right-to-left languages is not
yet provided. Speakers and on-camera actors are sourced from
publicly licensed media; a per-task demographic balance audit is
planned but not yet delivered. Out-of-scope by construction: live
streaming, real-time interactive media, very long-form content
(>1\,h), audio-less gameplay capture, and embodied / robotic-camera
footage.

\paragraph{Synthesis manifest.}
Where targeted synthetic inserts are required, all generation runs
through the following permissively licensed tools (full per-asset
provenance lives in \texttt{media.toml} and the Croissant
\texttt{rai:machineAnnotationTools} field):

\begin{table*}[h]
\centering
\small
\caption{Synthesis-tooling manifest. All tools listed are
permissively licensed; per-asset provenance is recorded in each
task's \texttt{media.toml}.}
\label{tab:appendix-synth-tools}
\begin{tabular}{ll}
\toprule
Tool & Use \\
\midrule
\texttt{ffmpeg} & audio/video processing, defect injection \\
FluidSynth + FluidR3\_GM SoundFont & MIDI $\to$ audio rendering \\
LilyPond 2.26.0 & music notation engraving (Bach, Mozart, etc.) \\
Kokoro-82M (Apache-2.0) & speech synthesis (10 distinct voices) \\
Godot 4.x + Kenney (CC0) & gameplay-QA footage rendering \\
Wav2Lip & lip-sync video synthesis (one task) \\
\texttt{reportlab} / \texttt{wkhtmltopdf} & PDF document synthesis \\
\texttt{matplotlib} & diagram rendering (educational tasks) \\
\texttt{music21} & MIDI extraction from public-domain scores \\
custom \texttt{build\_assets.py} & per-task deterministic synthesis scripts \\
\bottomrule
\end{tabular}
\end{table*}

\paragraph{Aggregate license breakdown.}
Per-asset license counts across the 497 source files in the
benchmark:

\begin{table*}[h]
\centering
\small
\caption{Aggregate license breakdown for source files across the
105-task suite. Counts derive from the per-asset license fields
recorded in each task's \texttt{media.toml}; non-commercial and
GPL-family files are addressed in the license caveats below.}
\label{tab:appendix-license-breakdown}
\begin{tabular}{lr l}
\toprule
License family & N files & Examples \\
\midrule
Apache-2.0 (incl.\ Kokoro-82M TTS) & 90 & TTS-synthesised speech \\
MIT & 78 & Author-synthesised assets \\
ODbL-1.0 & 88 & Map-tile derivatives \\
Public Domain / PD & 84 & Historical recordings, NASA \\
CC-BY family & 110 & Blender open movies, Wikimedia, author-contributed \\
CC0-1.0 (Kenney) & 35 & Game graphics + SFX \\
GPL family & 7 & Footage in \texttt{game-alert-mismatch} (see caveats) \\
CC-BY-NC-* & 3 & Three lecture-clip tasks (see caveats) \\
GFDL 1.2 & 2 & Wikimedia legacy \\
\bottomrule
\end{tabular}
\end{table*}

\paragraph{License caveats.}
Three lecture-content tasks (\texttt{chapter-repair},
\texttt{lecture-demo-clip-extract},
\texttt{long-form-clip-miner}) include MIT OpenCourseWare clips
under CC-BY-NC-SA-4.0 or CC-BY-NC-4.0; redistribution by users for
commercial purposes requires separate licensing. One game-QA task
(\texttt{game-alert-mismatch}) includes seven footage clips from
GPL-licensed open-source games (six under GPL-3.0+, one under
GPL-2.0+); downstream users redistributing that task's media bytes
are subject to the GPL's share-alike obligations on the bytes
themselves (the surrounding verifier and oracle code are
unaffected). Both license families permit academic-research use,
which is the benchmark's intended purpose. Per-file licenses are
recorded in each task's \texttt{media.toml}, and aggregate
\texttt{rai:dataLicense} entries in the Croissant metadata reflect
the same disclosure.

\subsection{Capability-Tag Co-Occurrence}

\begin{figure}[h]
\centering
\includegraphics[width=0.75\linewidth]{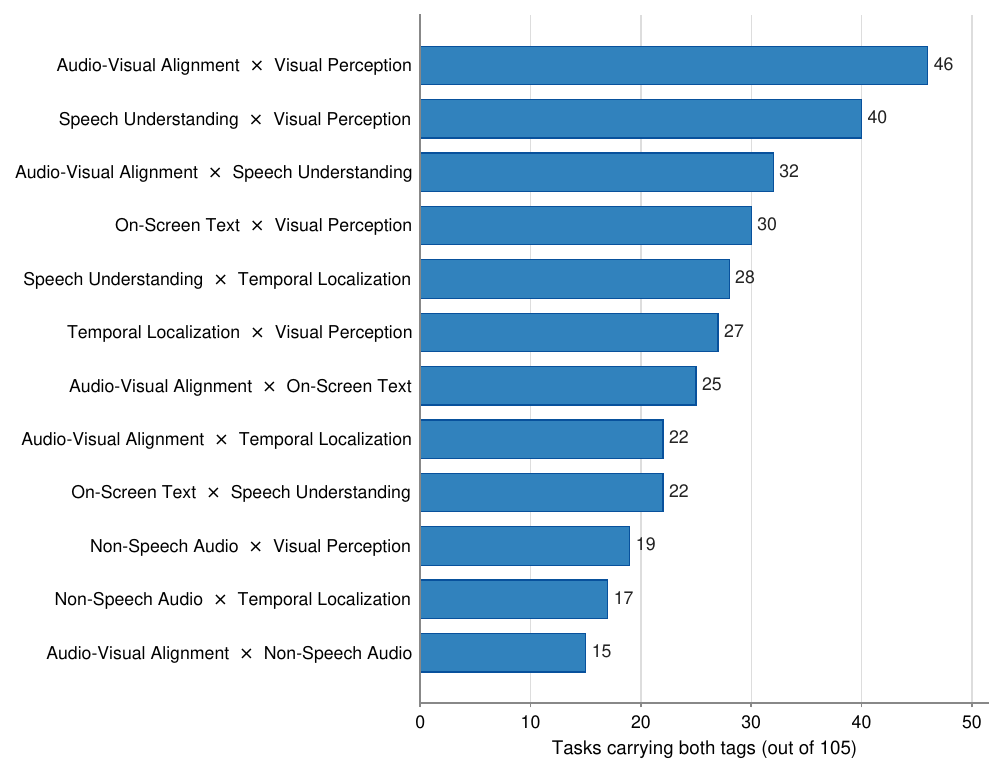}
\caption{Top capability-tag co-occurrence pairs across the
105-task suite under the twelve canonical capability tags. Each
bar counts the tasks that carry both tags. Strongest pairs are
Audio-Visual Alignment $\times$ Visual Perception (46 tasks),
Speech Understanding $\times$ Visual Perception (40), and
Audio-Visual Alignment $\times$ Speech Understanding (32),
reflecting the multimodal-grounding emphasis of the suite.}
\label{fig:tag-cooccurrence}
\end{figure}

\input{tex-table-figure/tag_radar_figure}

\input{tex-table-figure/capability_tag_breakdown.tex}

\input{tex-table-figure/category_breakdown.tex}

\subsection{Task Inventory by Fine-Grained Category}
\label{app:task-inventory}

The five meta-categories used throughout the paper (Table~\ref{tab:category-breakdown}, Table~\ref{tab:appendix-intake-counts}) are defined as follows.
\textbf{Media Production} ($n{=}40$): tasks that produce or quality-check media artifacts for an external audience, spanning subtitling and localization, broadcast and film post-production, podcast assembly, game-capture review, and social-media clip mining.
\textbf{Performance \& Coaching} ($n{=}23$): tasks that evaluate a third party's executed performance --- an actor's take, a language learner's pronunciation, or a musician's rendition --- and localize what was done correctly or incorrectly.
\textbf{Enterprise \& Compliance} ($n{=}23$): workplace tasks that extract decisions or audit evidence from meeting and screen-share recordings, or that structure business documents such as receipts, invoices, and reports.
\textbf{Personal \& Education} ($n{=}13$): tasks operating on the user's own media or learning materials, including personal and family recordings and lecture content consumed for self-study.
\textbf{Operations \& Research} ($n{=}6$): operational and research workflows without an external audience or performer, including ML dataset annotation, smart-device automation, and public-safety surveillance audits.

\begin{table*}[h]
\centering
\small
\caption{Implemented task counts by fine-grained category, grouped
by meta-category. All 16 canonical categories are populated;
totals sum to 105 done tasks across 5 meta-categories.}
\label{tab:appendix-intake-counts}
\begin{tabular}{llr}
\toprule
Meta-category & Fine-grained category & N \\
\midrule
\multirow{5}{*}{Media Production (40)}
  & Broadcast \& Film Production & 17 \\
  & Subtitling \& Localization & 8 \\
  & Game QA \& Esports & 6 \\
  & Audio Engineering \& Podcast Production & 6 \\
  & Creator Economy \& Social Media & 3 \\
\midrule
\multirow{3}{*}{Performance \& Coaching (23)}
  & Music Coaching \& Performance Feedback & 9 \\
  & Acting \& Casting & 8 \\
  & Language Learning \& Speech Coaching & 6 \\
\midrule
\multirow{3}{*}{Enterprise \& Compliance (23)}
  & Corporate Workflows \& Meetings & 13 \\
  & Compliance, Privacy \& Public Release & 6 \\
  & Document Processing \& Bookkeeping & 4 \\
\midrule
\multirow{2}{*}{Personal \& Education (13)}
  & Education \& Lecture Content & 8 \\
  & Personal / Everyday & 5 \\
\midrule
\multirow{3}{*}{Operations \& Research (6)}
  & Dataset \& ML Annotation & 4 \\
  & Automation \& Smart Devices & 1 \\
  & Public Safety \& City Ops & 1 \\
\midrule
\textbf{Total} & & \textbf{105} \\
\bottomrule
\end{tabular}
\end{table*}

\subsection{Per-Task Media Volume and Difficulty}
\label{app:media-volume}

We report per-task media-volume statistics as a proxy for the labor a
human practitioner would face on each task. All durations are
ffprobe-measured against the canonical \texttt{sha256}-pinned asset
files declared in \texttt{media.toml}; image and PDF entries
contribute to file counts but not to duration totals.

\begin{table*}[h]
\centering
\small
\setlength{\tabcolsep}{4pt}
\caption{Per-meta-category media-volume statistics (averages per task).
``files'' is the mean count of all asset files; ``image / video / audio''
break that count down by extension (other static formats such as PDF
contribute to the file count but are not enumerated). ``Avg.\ duration''
sums per-task video and audio duration and averages across the tasks
in the meta-category. The corpus totals 536 media files and 6\,h
54\,min of timed video and audio.}
\label{tab:appendix-media-volume}
\begin{tabular}{lrrrrrr}
\toprule
Meta-category & $n$ & files & image & video & audio & Avg.\ duration \\
\midrule
Media Production       & 40 & 4.08 & 0.07 & 2.90 & 0.82 & 5\,m\,20\,s \\
Performance \& Coaching & 23 & 6.13 & 0.39 & 0.35 & 4.91 & 0\,m\,52\,s \\
Enterprise \& Compliance & 23 & 4.35 & 0.26 & 0.96 & 1.78 & 1\,m\,45\,s \\
Personal \& Education  & 13 & 3.08 & 0.85 & 1.69 & 0.46 & 9\,m\,50\,s \\
Operations \& Research &  6 & 15.33 & 8.33 & 6.00 & 0.83 & 2\,m\,15\,s \\
\midrule
\textbf{All tasks}     & \textbf{105} & \textbf{5.10} & \textbf{0.75} & \textbf{1.94} & \textbf{1.89} & \textbf{3\,m\,57\,s} \\
\bottomrule
\end{tabular}
\end{table*}

The mix differs by meta-category: Performance \& Coaching is
audio-heavy (4.91 audio files / task on average, mostly short music
or speech samples), Personal \& Education concentrates timed content
in long-form lectures (9\,m\,50\,s mean per task), and Operations \&
Research carries the largest file counts per task (15.33 files,
mostly image batches for dataset annotation). The corpus-wide
distribution is also skewed: median per-task duration is 1\,m\,20\,s
while the mean is 3\,m\,57\,s and the maximum is 48\,m\,55\,s on the
longest single-lecture task.

\begin{figure}[h]
\centering
\includegraphics[width=\linewidth]{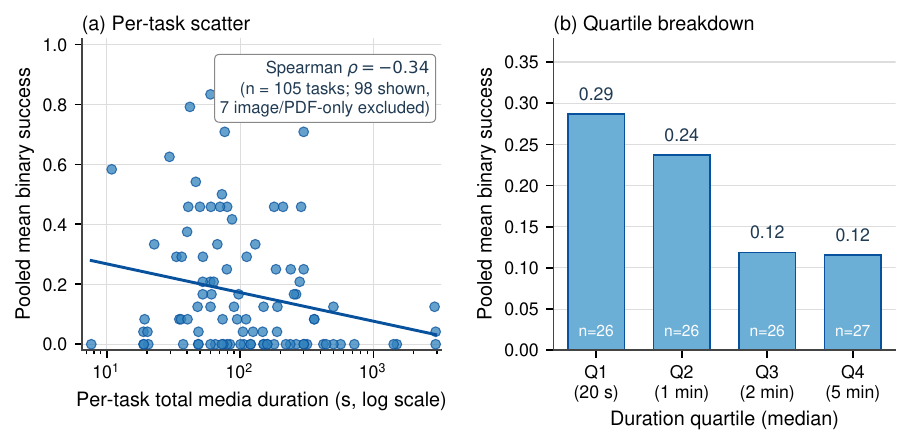}
\caption{Per-task media duration vs.\ pooled mean binary success
across all 11 evaluated harness$\times$model cells. (a) Scatter on
log-x with a least-squares fit (Spearman $\rho = -0.34$); the seven
image- or PDF-only tasks with zero timed media duration are shown in
panel (b)'s leftmost quartile but excluded from the log-x scatter.
(b) The same data binned into four equal-count duration quartiles:
mean binary success drops from 28.7\% in the shortest quartile
($\le 42$\,s) to 11.6\% in the longest quartile (median 5\,min,
range 210\,s--49\,min). The trend is monotonic but moderate; no
single duration threshold is the cause of the gap, and
file-count quartiles show no comparable trend
($\rho \approx 0$ for files-vs-success, not shown).}
\label{fig:media-volume-diagnostic}
\end{figure}

\subsection{Strategy-Divergence Case Studies}

\begin{figure}[h]
\centering
\includegraphics[width=\linewidth]{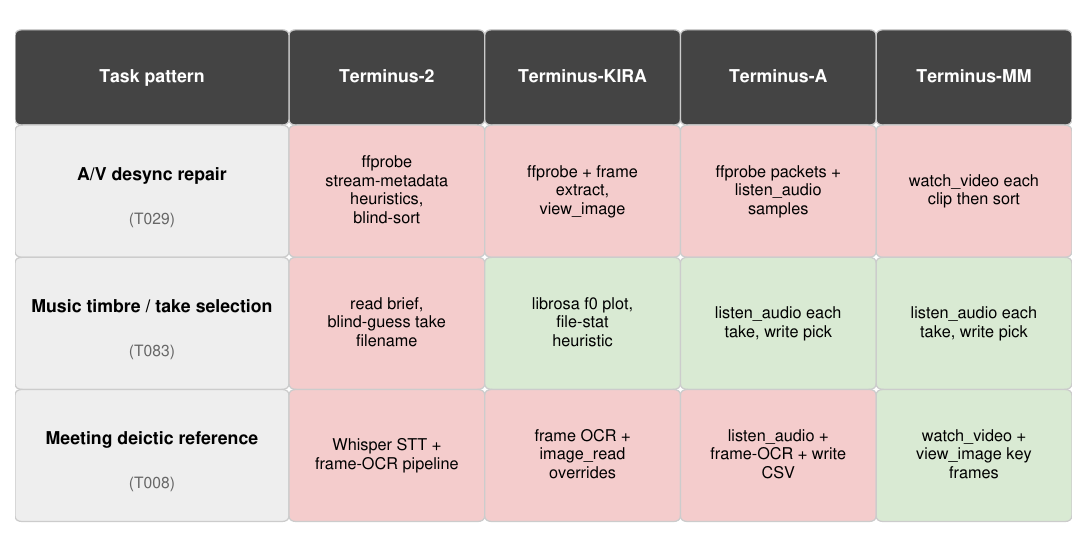}
\caption{
  Representative strategy divergence across harnesses on
  Gemini-3.1-Pro. Each cell summarizes the dominant per-turn
  pattern observed in the agent's trajectory; cell color marks
  whether that harness solved the task (green) or failed (red).
}
\label{fig:strategy-divergence}
\end{figure}

Figure~\ref{fig:strategy-divergence} illustrates the per-turn behaviour of
each Gemini-3.1-Pro harness on a small set of representative tasks, with
each cell summarising the dominant action pattern in the agent's
trajectory and coloured by whether that harness solved (green) or failed
(red) the task. The grid makes the strategy contrasts behind the
aggregate numbers in Section~\ref{sec:results-harness} concrete: harnesses
with native multimedia access tend to inspect the raw files directly,
while text-only harnesses route through command-line proxies, and the
visual divergence in trajectories matches the success/failure pattern
attributed to those access tiers in the main results.

\subsection{Domain-Level Modality Patterns (Heatmap)}

\begin{figure}[h]
\centering
\includegraphics[width=0.85\linewidth]{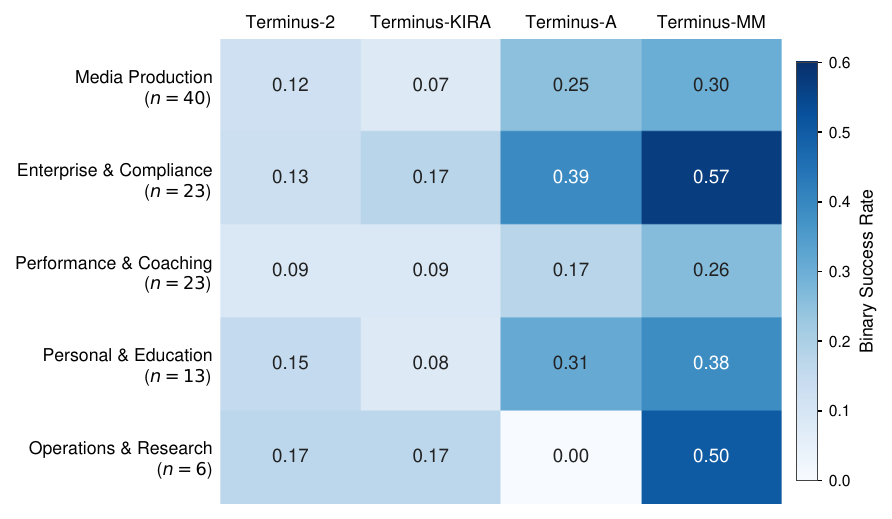}
\caption{
  Per-domain modality dependency on Gemini-3.1-Pro. Cells show
  mean binary success across the 105 tasks for each
  meta-category $\times$ harness combination. The four neutral
  harnesses span text only (T), text+image (KIRA), text+audio
  (A), and full multimodal routed (MM).
}
\label{fig:domain-heatmap}
\end{figure}

Figure~\ref{fig:domain-heatmap} presents mean binary success across the
105 tasks for each meta-category $\times$ harness combination on
Gemini-3.1-Pro, with rows spanning text-only (T), text+image (KIRA),
text+audio (A), and full multimedia routed (MM) harnesses. The heatmap
view shows that the modality-access gap reported in
Section~\ref{sec:results} is not uniform across workflows: it
concentrates in meta-categories where decisive evidence is audio-borne
or requires joint audio-visual grounding, whereas categories that admit
shell-based proxies show smaller gaps between MM and the partial-access
harnesses.

\subsection{Regime Analysis for \textsc{Terminus-MM} and Codex CLI}
\label{app:regime-analysis}

This appendix supports the solver-regime analysis in
Section~\ref{sec:results-harness}. We compare task-level binary outcomes
between \textsc{Terminus-MM} with Gemini-3.1-Pro and Codex CLI with GPT-5.2.
Here, \textsc{Terminus-MM} denotes the routed full-modality harness reported in
Table~\ref{tab:main-results}. The comparison is observational: the two systems
differ in both backbone and harness, so the partitions should be read as
observed solver regimes rather than causal proofs about which modality is
strictly necessary.

\paragraph{Partition construction.}
For each of the 105 tasks, we take the binary-pass outcome for
\textsc{Terminus-MM} and Codex CLI and assign the task to one of four regimes:
both systems pass, only Codex CLI passes, only \textsc{Terminus-MM} passes, or
both systems fail. All 105 tasks are paired; no task is missing from either
sweep. Table~\ref{tab:regime-partition} gives the resulting partition.

\begin{table*}[t]
\centering
\small
\caption{Task-level binary-pass partition for \textsc{Terminus-MM} and Codex CLI.}
\label{tab:regime-partition}
\begin{tabular}{@{}llrr@{}}
\toprule
Regime & Definition & Tasks & Share \\
\midrule
Both solve
& Codex pass $\wedge$ \textsc{MM} pass
& 11 & 10.5\% \\
Codex only
& Codex pass $\wedge$ \textsc{MM} fail
& 6 & 5.7\% \\
\textsc{MM} only
& Codex fail $\wedge$ \textsc{MM} pass
& 28 & 26.7\% \\
Both fail
& Codex fail $\wedge$ \textsc{MM} fail
& 60 & 57.1\% \\
\bottomrule
\end{tabular}
\end{table*}

\paragraph{Regime interpretation.}
The four-way split shows that the two systems are not related by strict
containment. The \textsc{MM}-only set is larger than the Codex-only set, but
the Codex-only set is non-empty and contains tasks where terminal pipelines
outperform the full-modality harness. We therefore interpret the regimes as
different bottlenecks. Codex-only tasks are \emph{pipeline-limited}: the media
evidence can be converted into stable intermediate signals and executable
edits. \textsc{MM}-only tasks are \emph{grounding-limited}: the output depends
on matching audio evidence to visual events, screen states, or temporally
localized actions. Both-fail tasks are \emph{combined-bottleneck} cases:
solving them requires both precise media grounding and robust terminal artifact
construction.

\begin{table*}[t]
\centering
\small
\caption{Metadata signatures of the four solver regimes.}
\label{tab:regime-metadata}
\begin{tabular}{@{}p{0.16\linewidth}p{0.30\linewidth}p{0.42\linewidth}@{}}
\toprule
Regime & Main metadata signals & Interpretation \\
\midrule
Both solve
& 11 tasks; lower native A/V requirement than the disagreement regimes; many JSON/CSV or simple edit outputs
& {Either system can reach the artifact once the media is reduced to reliable derived signals --- a transcript, a handful of frames, or a feature table --- through ordinary command-line tools.} \\
\addlinespace
Codex only
& 6 tasks; audio+video in 4/6; keywords cluster around OCR, DSP, synchronization, repair, and deterministic media editing
& The decisive evidence is accessible through {ordinary command-line operations --- transcribing speech, OCR'ing frames, computing signal features ---} so strong scripting and artifact repair can beat native perception. \\
\addlinespace
\textsc{MM} only
& 28 tasks; audio+video in 17/28; native audio in 23/28; CSV/JSON outputs in 67.8\%
& Grounding-limited workflows. Many outputs are per-event records where each row depends on co-grounding audio cues with visual events or states. \\
\addlinespace
Both fail
& 60 tasks; native audio in 59/60; video in 37/60; joint-A/V keyword in 21/60; JSON outputs in 68.3\%
& Remaining headroom. The hardest cases combine audio-rich evidence, joint A/V grounding, timing precision, and structured artifact production. \\
\bottomrule
\end{tabular}
\end{table*}

\paragraph{Task lists for the solved and disagreement regimes.}
The both-solve, Codex-only, and \textsc{MM}-only regimes are enumerated
below. The Codex-only tasks illustrate why the main text does not treat
native perception as a strict superset of terminal skill: their
deliverables can be recovered through OCR, timestamped transcription,
signal-energy analysis, synchronization repair, or deterministic file
edits. The \textsc{MM}-only tasks are enriched for per-event
audio-visual records: the output often requires deciding, for each row,
which spoken or sonic cue corresponds to which visual state, event, or
action.

\paragraph{Both-solve (11 tasks).}
\begin{multicols}{2}
\begin{itemize}[leftmargin=*,nosep]
  \setlength{\itemsep}{0pt}
  \item \texttt{broadcast-package-edit}
  \item \texttt{code-review-comment-attribution}
  \item \texttt{interview-srt-refine}
  \item \texttt{musical-mood-shot-pick}
  \item \texttt{narration-drift-qc}
  \item \texttt{near-duplicate-frame-dedup}
  \item \texttt{proof-step-note}
  \item \texttt{receipt-photo-to-json}
  \item \texttt{semantic-image-retrieval}
  \item \texttt{signal-based-qc-report}
  \item \texttt{warehouse-sku-pack-audit}
\end{itemize}
\end{multicols}

\paragraph{Codex-only (6 tasks).}
\begin{multicols}{2}
\begin{itemize}[leftmargin=*,nosep]
  \setlength{\itemsep}{0pt}
  \item \texttt{av-desync-offset-repair}
  \item \texttt{constant-hum-attenuation}
  \item \texttt{debate-attribution}
  \item \texttt{page-photo-to-text}
  \item \texttt{stereo-channel-flip-repair}
  \item \texttt{stream-alert-ack-audit}
\end{itemize}
\end{multicols}

\paragraph{\textsc{MM}-only (28 tasks).}
\begin{multicols}{2}
\begin{itemize}[leftmargin=*,nosep]
  \setlength{\itemsep}{0pt}
  \item \texttt{accessibility-sync-audit}
  \item \texttt{animation-narration-audit}
  \item \texttt{audience-ringtone-detection}
  \item \texttt{av-desync-detection}
  \item \texttt{blind-audition-match}
  \item \texttt{bug-repro-claim-audit}
  \item \texttt{caption-nonspeech-enrichment}
  \item \texttt{cooking-instruction-alignment}
  \item \texttt{crm-compliance-audit}
  \item \texttt{dead-air-removal}
  \item \texttt{design-review-approval-audit}
  \item \texttt{game-alert-mismatch}
  \item \texttt{interview-music-ducking-audit}
  \item \texttt{invoice-estimate-pdfs-to-xlsx}
  \item \texttt{lecturer-visual-term-ref}
  \item \texttt{line-failure-annotation}
  \item \texttt{lipsync-drift-correction}
  \item \texttt{narration-mars-rover}
  \item \texttt{narration-music-ducking}
  \item \texttt{partial-srt-resync}
  \item \texttt{prosody-multi-dim-selection}
  \item \texttt{prosody-take-selection}
  \item \texttt{screenshare-deictic-grounding}
  \item \texttt{slack-action-extraction}
  \item \texttt{speaker-action-attribution}
  \item \texttt{speaker-roster-identification}
  \item \texttt{tempo-drift-detection}
  \item \texttt{tutorial-edit-recreation}
\end{itemize}
\end{multicols}

\paragraph{Both-fail (60 tasks).}
The both-fail region contains the remaining 60 tasks. Representative
high-headroom cases include \texttt{narration-visual-align},
\texttt{2-speaker-diarized-transcript-from-podcast-audio},
\texttt{av-privacy-exposure}, and \texttt{multicam-active-speaker-cut}.
We summarize this region rather than listing all 60 names because its
main role in the paper is to characterize remaining benchmark headroom.

\paragraph{Representative trajectory evidence.}
Table~\ref{tab:regime-examples} gives one representative task from each regime.
The examples are not used to define the regimes; they are used to interpret why
the aggregate partition has the observed shape.

\begin{table*}[t]
\centering
\small
\setlength{\tabcolsep}{4pt}
\caption{Representative examples from the four solver regimes.}
\label{tab:regime-examples}
\begin{tabular}{@{}p{0.30\linewidth}ccp{0.46\linewidth}@{}}
\toprule
Task & Codex & \textsc{MM} & Interpretation \\
\midrule
\texttt{page-photo-to-text}
& 1.00 & 0.00
& Codex succeeds by constructing an OCR-centered document pipeline. The case illustrates that some media workflows are primarily pipeline-and-artifact problems rather than native-perception problems. \\
\addlinespace
\texttt{accessibility-sync-audit}
& 0.00 & 1.00
& The task requires aligning screen-reader audio with visual focus states and writing a row-level audit record. Codex {tries to inspect the audio--visual alignment by transcribing the audio and OCR'ing frames} but does not finish {in budget}; \textsc{MM} uses native A/V grounding to enumerate the events. \\
\addlinespace
\texttt{interview-srt-refine}
& 0.955 & 0.960
& Both systems reach near-perfect outputs. Their trajectories converge on similar transcript and boundary-refinement operations, {so the task is solvable from a transcript alone}. \\
\addlinespace
\texttt{2-speaker-\allowbreak diarized-\allowbreak transcript-\allowbreak from-podcast-audio}
& 0.00 & 0.582
& Both systems fall short of binary success. Codex spends the run attempting to set up diarization pipelines; \textsc{MM} identifies the speakers but lacks the required turn-boundary precision. \\
\bottomrule
\end{tabular}
\end{table*}

\paragraph{Synthesis.}
The regime analysis supports the interpretation in
Section~\ref{sec:results-harness}. Codex-only tasks show that strong terminal
agents can still win when the media can be reduced to reliable
{derived signals (a transcript, an OCR dump, or a
signal-feature table)} and the main challenge is executing the right pipeline. \textsc{MM}-only tasks
show that native media grounding is valuable when the artifact depends on
event-level correspondence between audio and visual evidence. The both-fail
region shows that the next frontier is not simply stronger perception or
stronger scripting in isolation, but agents that combine both: they must find
the decisive media evidence and complete the terminal artifact reliably under a
fixed interaction budget.

\subsection{{Filtering for the Cost Comparison in Section~\ref{sec:result-cost}}}
\label{app:proxy-cost}

This appendix supports the cost analysis in Section~\ref{sec:result-cost}{,
which compares the cost of inspecting a missing modality through command-line
tools to the cost of using native perception on the same task}.
All matched-pair rows are reported in Table~\ref{tab:proxy-cost}.

\paragraph{Filtering procedure.}
For each partial system, a task enters the {matched-pair} set
only if it passes
three filters: both the partial system and {Terminus-MM} pass the task
evaluator; the task requires a modality missing from the partial system; and
the partial trajectory {shows the agent attempting to inspect
the missing modality through command-line tools (extracted frames, transcripts,
OCR dumps, or signal-feature tables)}. Table~\ref{tab:proxy-filter-counts}
shows the filter
counts.

\begin{table*}[t]
\centering
\small
\caption{{\textbf{Matched-pair filter counts.} For each
partial system: how many of the 105 tasks both it and {Terminus-MM}
pass (Co-success), how many of those require a modality missing from the
partial system (Modality required), and how many of those show the agent
running command-line tools --- frame extraction, ASR, OCR, signal-feature
scripts --- to inspect the missing modality (Command-line attempts).}}
\label{tab:proxy-filter-counts}
\resizebox{\textwidth}{!}{%
\begin{tabular}{@{}llrrrr@{}}
\toprule
System & Missing access & Total & Co-success & Modality required & {Command-line attempts} \\
\midrule
\textsc{Terminus-2} (T)
    & image+audio+video & 105 & 8  & 8  & 7  \\
\textsc{KIRA} (T+I)
    & audio+video       & 105 & 8  & 5  & 4  \\
\textsc{A} (T+A)
    & image+video       & 105 & 21 & 18 & 14 \\
\textsc{V} (T+V)
    & image+audio       & 105 & 25 & 25 & 13 \\
\textsc{IA} (T+I+A)
    & video             & 105 & 28 & 18 & 13 \\
\textsc{IV} (T+I+V)
    & audio             & 105 & 29 & 23 & 14 \\
\textsc{AV} (T+A+V)
    & image             & 105 & 24 &  7 &  4 \\
Codex CLI (GPT-5.2, T+I)
    & image+audio+video & 105 & 11 & 11 & 11 \\
\bottomrule
\end{tabular}%
}
\end{table*}

\paragraph{{Where the overhead comes from.}}
The high-overhead cases share a small set of trajectory patterns. Partial
systems repeatedly {try to extract the missing} evidence by
{sampling} frames or audio
segments, running OCR or transcription, installing analysis libraries, computing
signal features, and iteratively refining timestamps or candidate artifacts.
These are valid terminal strategies, but they become costly when native
perception would provide the decisive evidence directly.

\subsection{Trajectory Evidence for Routed Perception-Tool Schemas}
\label{app:tool-exposure}

This appendix supports the analysis in
Section~\ref{sec:results-harness}. We provide implementation details for the
routed \textsc{Terminus-MM} harness and trajectory evidence showing why
unconditional perception-tool exposure can hurt full-modality agents.

\paragraph{Routed schema construction.}
\textsc{Terminus-MM w/o modality masking} exposes all native perception tools regardless of the
files present in the workspace. \textsc{Terminus-MM} instead constructs the
native perception-tool schema automatically at run start. Before the first LLM
call, the harness runs a bounded workspace file search and maps observed file
extensions to media types. Command execution and task completion tools are
always retained. The visual perception tool is retained when any media file is
present, since visual representations such as frames can be produced through
terminal media tools. The audio perception tool is retained only when an audio
file is present, and the video perception tool is retained only when a video
file is present.

This routing step is deterministic and task-uniform. It does not inspect the
task instruction, task identity, evaluator, reference solution, or answer. It
uses only filesystem information that the agent could also obtain through the
terminal. The routed harness therefore changes the perception-tool schema, not
the task, model, evaluator, or media assets.

\paragraph{Observed failure mode in \textsc{Terminus-MM w/o modality masking}.}
We identified the failure mode by inspecting \textsc{Terminus-MM w/o modality masking}
trajectories from the Gemini-3.1-Pro sweep. In video-only workspaces, the agent
sometimes first used native video perception, then created separate audio clips
from the video and invoked native audio perception on those derived files. This
second perception pass was not always harmless verification. In the inspected
cases, it either consumed the remaining interaction budget before the agent
wrote the required artifact, or pushed the agent toward an imprecise timing
commitment.

\begin{table*}[t]
\centering
\scriptsize
\setlength{\tabcolsep}{3pt}
\caption{
Failure mechanisms in inspected \textsc{Terminus-MM w/o modality masking} trajectories. All three
workspaces contain video files but no separate audio file.
}
\label{tab:tool-exposure-failure-cases}
\begin{tabular}{@{}p{0.20\linewidth}p{0.18\linewidth}p{0.26\linewidth}p{0.22\linewidth}c@{}}
\toprule
Task & Repeated media inspection & Failure mechanism & Terminal outcome & Reward \\
\midrule
\texttt{spoken-decision-\allowbreak cell-ref}
& 2 native audio calls on derived clips after native video perception
& Time exhaustion after repeated audio re-checking and visual micro-refinement
& No \texttt{notes.csv} written; task is not completed
& 0.00 \\
\addlinespace
\texttt{narration-\allowbreak visual-align}
& 6 native audio calls on derived segments after native video perception
& Time exhaustion before the agent reaches the JSON artifact write
& No \texttt{mismatches.json} written; task is not completed
& 0.00 \\
\addlinespace
\texttt{constant-\allowbreak offset-srt}
& 3 native audio calls on increasingly narrow derived clips
& Wrong commitment: transcript-level timing estimate leads to a uniform 700ms offset error
& \texttt{subs\_corrected.vtt} written, but shifted to the wrong boundary
& 0.64 \\
\bottomrule
\end{tabular}
\end{table*}

The key point is that repeated checking is not intrinsically wrong; it becomes
harmful when it delays or distorts the artifact-producing part of the workflow.
For \texttt{spoken-decision-cell-ref} and
\texttt{narration-visual-align}, the agent never writes the required output
file. For \texttt{constant-offset-srt}, the agent does write the output, but the
committed subtitle offset is uniformly 700ms away from the gold boundary. Under
the task evaluator's linear timing reward, this produces the observed partial
score of 0.64.

\paragraph{Effect of routing on the same cases.}
The routed \textsc{Terminus-MM} schema removes the audio perception tool from
these video-only workspaces because no separate audio file is present at run
start. The agent still has access to native video perception and ordinary
terminal media-processing tools, but it is not presented with native audio as a
separate verification primitive. On the three inspected cases, this removes the
repeated native-audio detour and the agent reaches task completion in all three
runs.

\begin{table*}[t]
\centering
\scriptsize
\setlength{\tabcolsep}{3pt}
\caption{
Task-level recoveries under routed \textsc{Terminus-MM} on the inspected
Gemini-3.1-Pro cases.
}
\label{tab:tool-exposure-recoveries}
\begin{tabular}{@{}p{0.22\linewidth}p{0.13\linewidth}p{0.10\linewidth}p{0.13\linewidth}p{0.32\linewidth}@{}}
\toprule
Task & MM w/o mask reward & MM reward & Native audio calls (routed) & Routed behavior \\
\midrule
\texttt{spoken-decision-\allowbreak cell-ref}
& 0.00 & 1.00 & 0
& Completes the required record instead of spending the final budget on repeated audio verification. \\
\addlinespace
\texttt{narration-\allowbreak visual-align}
& 0.00 & 0.60 & 0
& Reaches artifact production after using video perception and terminal-side processing. \\
\addlinespace
\texttt{constant-\allowbreak offset-srt}
& 0.64 & 0.96 & 0
& Avoids transcript-level boundary commitment and obtains a more accurate timing correction. \\
\bottomrule
\end{tabular}
\end{table*}

These trajectories explain the aggregate result in
Table~\ref{tab:tool-exposure}. \textsc{Terminus-MM} does not add a new
perception capability over \textsc{Terminus-MM w/o modality masking}; it removes perception tools
whose target file types are absent from the initial workspace. The improvement
therefore supports the interpretation in Section~\ref{sec:results-harness}: the
original full-modality harness was not only limited by media understanding, but
also by schema-induced tool-use behavior that could divert the agent away from
artifact completion.

%% file: tex-table-figure/modality_table_full.tex
\begin{table*}[t]
\centering
\small
\caption{
Modality ablation across backbones. Each Terminus row exposes a different
subset of perception tools to the agent: T = text-only shell;
I = \texttt{view\_image}; A = \texttt{listen\_audio}; V = \texttt{watch\_video}.
\textsc{Terminus-MM}'s schema is workspace-routed --- it drops perception tools
whose target file types are absent from the workspace at run start
(Section~\ref{sec:results-harness}); the remaining Terminus rows use a static
schema. Outcomes report mean binary success and mean task-specific partial
credit. Cost columns: API = mean USD per task; Tokens = mean total tokens
per task ($\text{input} + \text{cache} + \text{output}$, in thousands);
Time = mean agent execution wall (excludes container setup and verifier scoring).
}
\label{tab:main-results-full}
\resizebox{\textwidth}{!}{%
\begin{tabular}{@{}lllcc|ccc@{}}
\toprule
\multirow{2}{*}{Harness} &
\multirow{2}{*}{Model} &
\multirow{2}{*}{Access} &
\multicolumn{2}{c|}{Success Rate $\uparrow$} &
\multicolumn{3}{c}{Cost $\downarrow$} \\
& & & Binary & Partial & API & Tokens & Time \\
\midrule
Terminus-2    & \multirow{2}{*}{Qwen3.5-122B} & T   & 0.105 & 0.159 & \$0.101 & 308.3k & 510s \\
Terminus-KIRA &                               & T+I & 0.095 & 0.165 & \$0.233 & 725.8k & 519s \\
\midrule
Terminus-2    & \multirow{2}{*}{GPT-5.2}      & T   & 0.105 & 0.149 & \$0.818 & 153.5k & 500s \\
Terminus-KIRA &                               & T+I & 0.114 & 0.150 & \$1.672 & 248.6k & 540s \\
\midrule
Terminus-2     & \multirow{8}{*}{Gemini-2.5-Flash} & T        & 0.067 & 0.136 & \$0.115 & 503.6k & 248s \\
Terminus-KIRA  &                                    & T+I      & 0.067 & 0.129 & \$0.226 & 657.4k & 290s \\
Terminus-A     &                                    & T+A      & 0.105 & 0.183 & \$0.127 & 359.7k & 209s \\
Terminus-V     &                                    & T+V      & 0.229 & 0.307 & \$0.140 & 411.8k & 210s \\
Terminus-IA    &                                    & T+I+A    & 0.133 & 0.181 & \$0.234 & 706.1k & 269s \\
Terminus-IV    &                                    & T+I+V    & 0.162 & 0.222 & \$0.184 & 546.5k & 272s \\
Terminus-AV    &                                    & T+A+V    & 0.219 & 0.315 & \$0.139 & 406.8k & 188s \\
Terminus-MM    &                                    & T+I+A+V  & 0.229 & 0.305 & \$0.099 & 282.0k & 229s \\
\midrule
Terminus-2     & \multirow{8}{*}{Gemini-3.1-Pro} & T        & 0.124 & 0.162 & \$0.772 & 419.7k & 538s \\
Terminus-KIRA  &                                  & T+I      & 0.105 & 0.159 & \$2.061 & 932.4k & 544s \\
Terminus-A     &                                  & T+A      & 0.257 & 0.349 & \$1.992 & 904.9k & 493s \\
Terminus-V     &                                  & T+V      & 0.286 & 0.395 & \$1.118 & 481.5k & 417s \\
Terminus-IA    &                                  & T+I+A    & 0.333 & 0.406 & \$1.742 & 798.2k & 460s \\
Terminus-IV    &                                  & T+I+V    & 0.333 & 0.432 & \$1.283 & 560.7k & 434s \\
Terminus-AV    &                                  & T+A+V    & 0.276 & 0.408 & \$1.029 & 442.1k & 419s \\
Terminus-MM    &                                  & T+I+A+V  & 0.371 & 0.469 & \$1.228 & 538.1k & 442s \\
\midrule
Claude Code    & Sonnet-4.6 & T+I & 0.162 & 0.186 & \$1.735 &   981.1k & 516s \\
Codex CLI      & GPT-5.2    & T+I & 0.162 & 0.202 & \$7.117 & 2{,}620.6k & 529s \\
\bottomrule
\end{tabular}%
}
\end{table*}

%% file: tex-table-figure/tag_radar_figure.tex
\begin{figure}[t]
  \centering
  \includegraphics[width=\linewidth]{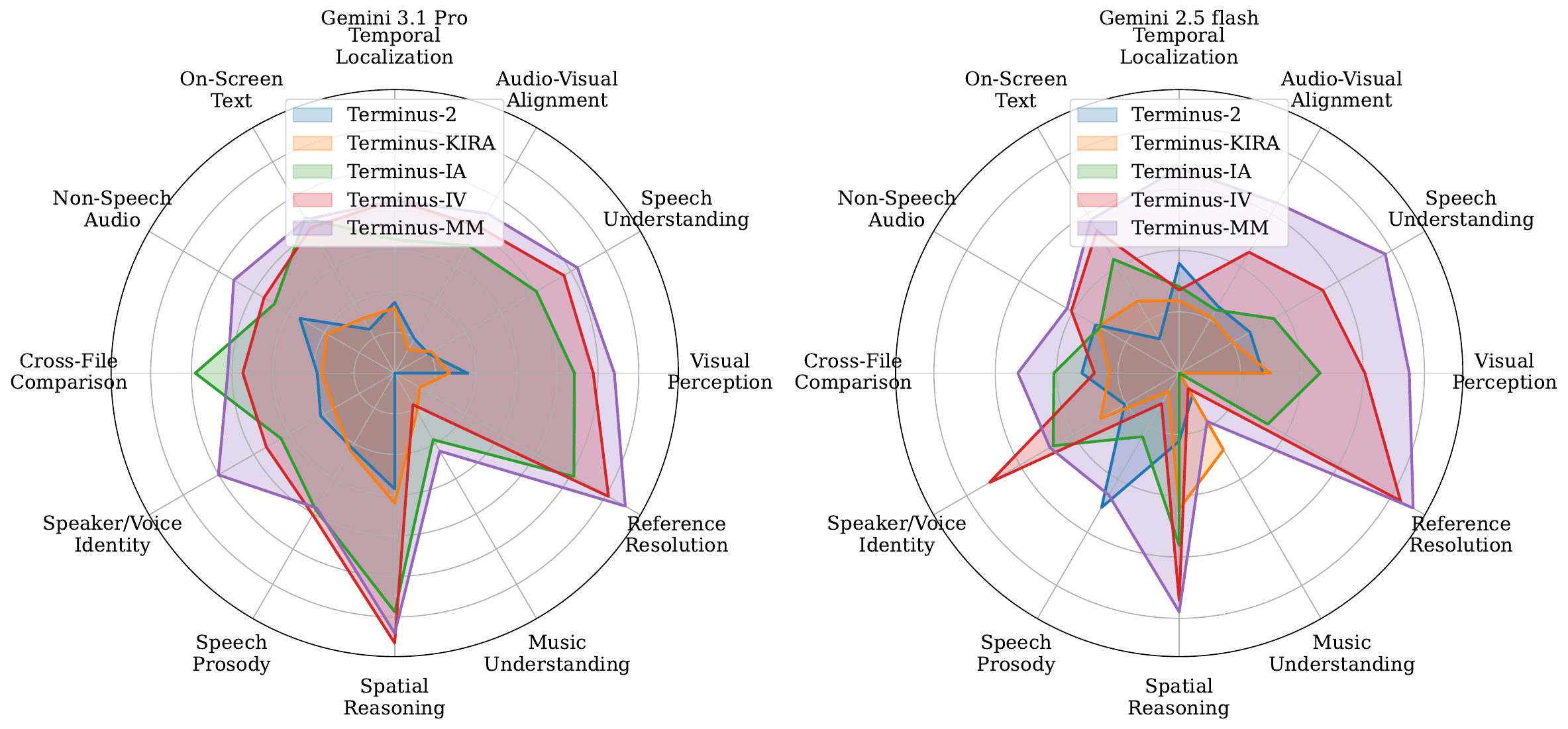}
  \caption{
    Partial Success rate in each capability tag.
  }
  \label{fig:tag-radar}
\end{figure}

%% file: tex-table-figure/capability_tag_breakdown.tex
% Per-capability-tag outcome/cost breakdown — auto-generated.
% Bold = best within Flash or Pro Terminus family; underline = 2nd-best within same family.
% \cellcolor{rankgold} = global #1 across all 16 backbone×harness cells per metric column;
% \cellcolor{ranksilver} = global #2.
%
% arXiv split: the original wide 36-column variant overflowed the page; this
% file now emits two stacked half-tables, each carrying 6 capability tags.
% Both halves share the same row structure (16 backbone×harness cells) and
% should be read together; cross-half aggregation is preserved by the legend.

\begin{table*}[!htbp]
\centering
\tiny
\setlength{\tabcolsep}{2pt}
\caption{Per-capability-tag success-rate and cost breakdown for all main-table backbone $\times$ harness cells (\textbf{Part 1 of 2}: tags 1--6). Capability tags are \emph{multi-label}: a task carrying $k$ tags contributes to all $k$ columns. Sum of $n$ across all 12 tag columns (Parts 1 \& 2 combined) is $365$ ($102$ of $105$ tasks carry $\geq 2$ tags; max $6$ tags/task). Per-cell metrics are unweighted means over the tasks bearing that tag. Per-tag column triplet reports binary success rate ($B$), partial success rate ($P$), and mean USD cost ($\$$). \textbf{Bold} marks the best harness within the Flash or Pro Terminus family (5 cells per family); \underline{underline} marks the 2nd-best within the same family. \colorbox{rankgold}{Gold} highlights the global \#1 cell across all 16 backbone$\times$harness rows for each metric column; \colorbox{ranksilver}{silver} highlights the global \#2.}
\label{tab:capability-tag-breakdown}
\resizebox{\textwidth}{!}{%
\begin{tabular}{@{}llcccccccccccccccccccc@{}}
\toprule
 &  & \multicolumn{3}{c}{Audio-Visual Alignment ($n{=}55$)} & \multicolumn{3}{c}{Cross-File Comparison ($n{=}21$)} & \multicolumn{3}{c}{Music Understanding ($n{=}11$)} & \multicolumn{3}{c}{Non-Speech Audio ($n{=}26$)} & \multicolumn{3}{c}{On-Screen Text ($n{=}42$)} & \multicolumn{3}{c}{Reference Resolution ($n{=}7$)} \\
\cmidrule(lr){3-5}\cmidrule(lr){6-8}\cmidrule(lr){9-11}\cmidrule(lr){12-14}\cmidrule(lr){15-17}\cmidrule(lr){18-20}
Backbone & Harness & $B$ & $P$ & \$ & $B$ & $P$ & \$ & $B$ & $P$ & \$ & $B$ & $P$ & \$ & $B$ & $P$ & \$ & $B$ & $P$ & \$ \\
\midrule
\multirow{2}{*}{Qwen3.5-122B} & \textsc{Terminus-2} & 0.036 & 0.062 & \$0.113 & 0.191 & 0.189 & \cellcolor{ranksilver}\$0.098 & 0.000 & 0.000 & \$0.158 & 0.115 & 0.187 & \cellcolor{ranksilver}\$0.116 & 0.095 & 0.118 & \cellcolor{ranksilver}\$0.112 & 0.000 & 0.000 & \$0.088 \\
 & \textsc{Terminus-KIRA} & 0.054 & 0.126 & \$0.244 & 0.095 & 0.121 & \$0.206 & 0.000 & 0.057 & \$0.199 & 0.115 & 0.211 & \$0.257 & 0.119 & 0.170 & \$0.232 & 0.000 & 0.000 & \$0.234 \\
\cmidrule(lr){1-20}
\multirow{2}{*}{GPT-5.2} & \textsc{Terminus-2} & 0.073 & 0.099 & \$0.813 & 0.238 & 0.275 & \$0.990 & \cellcolor{ranksilver}0.091 & 0.091 & \$0.940 & 0.115 & 0.202 & \$0.839 & 0.071 & 0.076 & \$0.806 & 0.000 & 0.000 & \$0.751 \\
 & \textsc{Terminus-KIRA} & 0.073 & 0.082 & \$1.747 & 0.048 & 0.067 & \$1.821 & 0.000 & 0.015 & \$1.395 & 0.154 & 0.209 & \$1.609 & 0.119 & 0.122 & \$1.799 & 0.143 & 0.143 & \$1.966 \\
\cmidrule(lr){1-20}
\multirow{5}{*}{Gemini-2.5-Flash} & \textsc{Terminus-2} & 0.091 & 0.126 & \cellcolor{rankgold}\textbf{\$0.097} & 0.143 & 0.159 & \underline{\$0.099} & 0.000 & 0.043 & \cellcolor{ranksilver}\underline{\$0.155} & 0.038 & 0.158 & \cellcolor{rankgold}\textbf{\$0.083} & 0.000 & 0.065 & \underline{\$0.172} & 0.000 & 0.000 & \$0.076 \\
 & \textsc{Terminus-KIRA} & 0.054 & 0.105 & \$0.280 & 0.048 & 0.114 & \$0.127 & \textbf{0.091} & \textbf{0.144} & \$0.213 & 0.077 & 0.152 & \$0.181 & 0.071 & 0.136 & \$0.300 & 0.000 & 0.000 & \$0.322 \\
 & \textsc{Terminus-IA} & 0.054 & 0.119 & \$0.276 & \underline{0.191} & \underline{0.204} & \$0.159 & 0.000 & 0.000 & \$0.261 & 0.077 & 0.149 & \$0.223 & \underline{0.191} & 0.214 & \$0.334 & 0.143 & 0.167 & \$0.050 \\
 & \textsc{Terminus-IV} & \underline{0.182} & \underline{0.228} & \$0.187 & 0.095 & 0.138 & \$0.166 & 0.000 & 0.029 & \$0.205 & \textbf{0.115} & \underline{0.203} & \underline{\$0.135} & 0.191 & \underline{0.269} & \$0.182 & \underline{0.286} & \underline{0.417} & \cellcolor{rankgold}\textbf{\$0.015} \\
 & \textsc{Terminus-MM} & \textbf{0.255} & \textbf{0.320} & \cellcolor{ranksilver}\underline{\$0.102} & \textbf{0.238} & \textbf{0.263} & \cellcolor{rankgold}\textbf{\$0.070} & \underline{0.091} & \underline{0.091} & \cellcolor{rankgold}\textbf{\$0.096} & \underline{0.115} & \textbf{0.211} & \$0.143 & \textbf{0.262} & \textbf{0.289} & \cellcolor{rankgold}\textbf{\$0.093} & \textbf{0.429} & \textbf{0.441} & \cellcolor{ranksilver}\underline{\$0.020} \\
\cmidrule(lr){1-20}
\multirow{5}{*}{Gemini-3.1-Pro} & \textsc{Terminus-2} & 0.091 & 0.097 & \textbf{\$0.779} & 0.191 & 0.191 & \underline{\$0.818} & 0.000 & 0.000 & \$1.072 & 0.231 & 0.269 & \textbf{\$0.859} & 0.119 & 0.124 & \textbf{\$0.741} & 0.000 & 0.000 & \textbf{\$0.540} \\
 & \textsc{Terminus-KIRA} & 0.054 & 0.067 & \$2.176 & 0.143 & 0.179 & \$2.106 & \underline{0.091} & 0.109 & \textbf{\$1.019} & 0.115 & 0.192 & \$2.308 & 0.143 & 0.156 & \$1.990 & 0.000 & 0.071 & \$2.233 \\
 & \textsc{Terminus-IA} & \cellcolor{ranksilver}\underline{0.291} & 0.362 & \$2.261 & \cellcolor{rankgold}\textbf{0.476} & \cellcolor{rankgold}\textbf{0.490} & \$1.385 & 0.091 & \cellcolor{ranksilver}\underline{0.189} & \underline{\$1.030} & \cellcolor{ranksilver}\underline{0.269} & 0.341 & \$2.205 & \cellcolor{rankgold}\textbf{0.357} & \cellcolor{rankgold}\textbf{0.440} & \$1.818 & 0.429 & 0.509 & \$2.111 \\
 & \textsc{Terminus-IV} & 0.255 & \cellcolor{ranksilver}\underline{0.415} & \underline{\$1.314} & 0.333 & 0.374 & \$0.927 & 0.000 & 0.089 & \$1.396 & 0.269 & \cellcolor{ranksilver}\underline{0.371} & \$1.311 & 0.333 & 0.413 & \underline{\$1.314} & \cellcolor{rankgold}\textbf{0.571} & \cellcolor{ranksilver}\underline{0.607} & \underline{\$1.005} \\
 & \textsc{Terminus-MM} & \cellcolor{rankgold}\textbf{0.327} & \cellcolor{rankgold}\textbf{0.454} & \$1.336 & \cellcolor{ranksilver}\underline{0.381} & \cellcolor{ranksilver}\underline{0.411} & \textbf{\$0.663} & \cellcolor{rankgold}\textbf{0.182} & \cellcolor{rankgold}\textbf{0.222} & \$1.256 & \cellcolor{rankgold}\textbf{0.385} & \cellcolor{rankgold}\textbf{0.457} & \underline{\$1.031} & \cellcolor{ranksilver}\underline{0.357} & \cellcolor{ranksilver}\underline{0.436} & \$1.351 & \cellcolor{ranksilver}\underline{0.571} & \cellcolor{rankgold}\textbf{0.655} & \$1.028 \\
\cmidrule(lr){1-20}
Sonnet-4.6 & \textsc{Claude Code} & 0.127 & 0.122 & \$1.877 & 0.286 & 0.287 & \$1.395 & 0.091 & 0.126 & \$1.523 & 0.269 & 0.337 & \$1.629 & 0.167 & 0.173 & \$1.868 & 0.143 & 0.143 & \$3.161 \\
\cmidrule(lr){1-20}
GPT-5.2 & \textsc{Codex CLI} & 0.109 & 0.139 & \$7.462 & 0.191 & 0.191 & \$5.755 & 0.091 & 0.091 & \$8.010 & 0.192 & 0.235 & \$5.958 & 0.167 & 0.218 & \$7.925 & 0.143 & 0.241 & \$11.920 \\
\bottomrule
\end{tabular}%
}
\end{table*}

\begin{table*}[!htbp]
\centering
\tiny
\setlength{\tabcolsep}{2pt}
\caption{Per-capability-tag success-rate and cost breakdown, \textbf{Part 2 of 2}: tags 7--12. Same structure and legend as Table~\ref{tab:capability-tag-breakdown}. The two halves together cover all 12 capability tags; readers should join Parts 1 and 2 along the row axis (the same 16 backbone$\times$harness rows appear in both).}
\label{tab:capability-tag-breakdown-part2}
\resizebox{\textwidth}{!}{%
\begin{tabular}{@{}llcccccccccccccccccccc@{}}
\toprule
 &  & \multicolumn{3}{c}{Spatial Reasoning ($n{=}12$)} & \multicolumn{3}{c}{Speaker/Voice Identity ($n{=}13$)} & \multicolumn{3}{c}{Speech Prosody ($n{=}13$)} & \multicolumn{3}{c}{Speech Understanding ($n{=}57$)} & \multicolumn{3}{c}{Temporal Localization ($n{=}44$)} & \multicolumn{3}{c}{Visual Perception ($n{=}68$)} \\
\cmidrule(lr){3-5}\cmidrule(lr){6-8}\cmidrule(lr){9-11}\cmidrule(lr){12-14}\cmidrule(lr){15-17}\cmidrule(lr){18-20}
Backbone & Harness & $B$ & $P$ & \$ & $B$ & $P$ & \$ & $B$ & $P$ & \$ & $B$ & $P$ & \$ & $B$ & $P$ & \$ & $B$ & $P$ & \$ \\
\midrule
\multirow{2}{*}{Qwen3.5-122B} & \textsc{Terminus-2} & 0.167 & 0.202 & \cellcolor{ranksilver}\$0.139 & 0.077 & 0.171 & \$0.134 & 0.154 & 0.280 & \$0.069 & 0.088 & 0.140 & \cellcolor{ranksilver}\$0.083 & 0.114 & 0.195 & \cellcolor{rankgold}\$0.089 & 0.088 & 0.137 & \cellcolor{ranksilver}\$0.108 \\
 & \textsc{Terminus-KIRA} & 0.333 & 0.352 & \$0.256 & 0.000 & 0.143 & \$0.184 & 0.077 & 0.138 & \$0.248 & 0.035 & 0.096 & \$0.231 & 0.114 & 0.174 & \$0.255 & 0.118 & 0.195 & \$0.248 \\
\cmidrule(lr){1-20}
\multirow{2}{*}{GPT-5.2} & \textsc{Terminus-2} & 0.083 & 0.110 & \$0.673 & 0.077 & 0.138 & \$0.924 & 0.077 & 0.171 & \$1.107 & 0.053 & 0.114 & \$0.719 & 0.114 & 0.170 & \$0.759 & 0.118 & 0.150 & \$0.833 \\
 & \textsc{Terminus-KIRA} & 0.167 & 0.191 & \$1.591 & 0.077 & 0.133 & \$1.472 & 0.000 & 0.090 & \$1.750 & 0.105 & 0.127 & \$1.786 & 0.182 & 0.213 & \$1.600 & 0.147 & 0.151 & \$1.782 \\
\cmidrule(lr){1-20}
\multirow{5}{*}{Gemini-2.5-Flash} & \textsc{Terminus-2} & 0.083 & 0.111 & \underline{\$0.172} & 0.077 & 0.103 & \cellcolor{rankgold}\textbf{\$0.086} & \textbf{0.154} & \textbf{0.253} & \cellcolor{rankgold}\textbf{\$0.044} & 0.070 & 0.133 & \underline{\$0.113} & 0.068 & \underline{0.179} & \cellcolor{ranksilver}\textbf{\$0.137} & 0.073 & 0.137 & \underline{\$0.114} \\
 & \textsc{Terminus-KIRA} & 0.167 & 0.222 & \$0.242 & 0.077 & 0.148 & \$0.233 & 0.000 & 0.035 & \$0.133 & 0.035 & 0.100 & \$0.282 & 0.068 & 0.118 & \$0.261 & 0.088 & 0.149 & \$0.240 \\
 & \textsc{Terminus-IA} & \underline{0.250} & 0.281 & \$0.205 & \underline{0.154} & 0.237 & \$0.145 & \underline{0.077} & 0.120 & \cellcolor{ranksilver}\underline{\$0.059} & 0.123 & 0.178 & \$0.237 & 0.091 & 0.141 & \$0.324 & 0.176 & 0.230 & \$0.232 \\
 & \textsc{Terminus-IV} & 0.167 & \underline{0.371} & \$0.261 & \cellcolor{rankgold}\textbf{0.308} & \textbf{0.357} & \$0.180 & 0.000 & 0.058 & \$0.191 & \underline{0.228} & \underline{0.271} & \$0.146 & \underline{0.114} & 0.135 & \$0.190 & \underline{0.235} & \underline{0.303} & \$0.196 \\
 & \textsc{Terminus-MM} & \textbf{0.333} & \textbf{0.390} & \cellcolor{rankgold}\textbf{\$0.061} & 0.154 & \underline{0.243} & \cellcolor{ranksilver}\underline{\$0.130} & 0.077 & \underline{0.230} & \$0.073 & \textbf{0.316} & \textbf{0.388} & \cellcolor{rankgold}\textbf{\$0.079} & \textbf{0.250} & \textbf{0.335} & \underline{\$0.140} & \textbf{0.309} & \textbf{0.375} & \cellcolor{rankgold}\textbf{\$0.089} \\
\cmidrule(lr){1-20}
\multirow{5}{*}{Gemini-3.1-Pro} & \textsc{Terminus-2} & 0.250 & 0.285 & \textbf{\$0.659} & 0.154 & 0.210 & \textbf{\$0.805} & 0.077 & 0.212 & \$1.020 & 0.053 & 0.096 & \textbf{\$0.686} & 0.159 & 0.174 & \textbf{\$0.727} & 0.162 & 0.179 & \textbf{\$0.685} \\
 & \textsc{Terminus-KIRA} & 0.250 & 0.321 & \$1.538 & 0.077 & 0.173 & \$1.576 & 0.077 & 0.220 & \$2.226 & 0.035 & 0.105 & \$2.335 & 0.114 & 0.160 & \$2.214 & 0.118 & 0.135 & \$2.178 \\
 & \textsc{Terminus-IA} & \cellcolor{ranksilver}\underline{0.500} & 0.588 & \$1.961 & \underline{0.231} & 0.322 & \underline{\$1.380} & \cellcolor{rankgold}\textbf{0.385} & \cellcolor{ranksilver}\underline{0.388} & \textbf{\$0.780} & 0.298 & 0.402 & \$1.943 & 0.250 & 0.329 & \$1.944 & \cellcolor{ranksilver}\underline{0.382} & 0.442 & \$2.081 \\
 & \textsc{Terminus-IV} & \cellcolor{rankgold}\textbf{0.583} & \cellcolor{rankgold}\textbf{0.664} & \underline{\$0.955} & 0.231 & \cellcolor{ranksilver}\underline{0.364} & \$1.703 & \cellcolor{ranksilver}\underline{0.308} & \cellcolor{rankgold}\textbf{0.403} & \$1.024 & \cellcolor{ranksilver}\underline{0.368} & \cellcolor{ranksilver}\underline{0.481} & \underline{\$1.302} & \cellcolor{ranksilver}\underline{0.341} & \cellcolor{rankgold}\textbf{0.426} & \$1.440 & 0.368 & \cellcolor{ranksilver}\underline{0.488} & \$1.215 \\
 & \textsc{Terminus-MM} & 0.500 & \cellcolor{ranksilver}\underline{0.640} & \$1.026 & \cellcolor{ranksilver}\textbf{0.308} & \cellcolor{rankgold}\textbf{0.501} & \$1.687 & 0.308 & 0.383 & \underline{\$0.906} & \cellcolor{rankgold}\textbf{0.386} & \cellcolor{rankgold}\textbf{0.519} & \$1.471 & \cellcolor{rankgold}\textbf{0.364} & \cellcolor{ranksilver}\underline{0.425} & \underline{\$1.334} & \cellcolor{rankgold}\textbf{0.441} & \cellcolor{rankgold}\textbf{0.540} & \underline{\$1.185} \\
\cmidrule(lr){1-20}
Sonnet-4.6 & \textsc{Claude Code} & 0.417 & 0.402 & \$1.710 & 0.154 & 0.151 & \$1.954 & 0.077 & 0.115 & \$1.315 & 0.088 & 0.097 & \$2.005 & 0.091 & 0.127 & \$1.774 & 0.235 & 0.236 & \$1.957 \\
\cmidrule(lr){1-20}
GPT-5.2 & \textsc{Codex CLI} & 0.250 & 0.307 & \$8.049 & 0.154 & 0.154 & \$5.993 & 0.000 & 0.038 & \$5.191 & 0.123 & 0.173 & \$8.064 & 0.159 & 0.211 & \$6.472 & 0.221 & 0.259 & \$7.721 \\
\bottomrule
\end{tabular}%
}
\end{table*}

%% file: tex-table-figure/category_breakdown.tex
% Per-meta-category outcome/cost breakdown — auto-generated.
% Bold = best within Flash or Pro Terminus family; underline = 2nd-best within same family.
% \cellcolor{rankgold} = global #1 across all 16 backbone×harness cells per metric column;
% \cellcolor{ranksilver} = global #2.
\begin{table*}[!htbp]
\centering
\scriptsize
\setlength{\tabcolsep}{4pt}
\caption{Per-meta-category success-rate and cost breakdown for all main-table backbone $\times$ harness cells. Five meta-categories cover all 105 tasks ($\sum n_{\text{tasks}} = 105$). Per-cell metrics are weighted means over the constituent fine-categories. Each meta-category column triplet reports binary success rate ($B$), partial success rate ($P$), and mean USD cost ($\$$). \textbf{Bold} marks the best harness within the Flash or Pro Terminus family (5 cells per family); \underline{underline} marks the 2nd-best within the same family. \colorbox{rankgold}{Gold} highlights the global \#1 cell across all 16 backbone$\times$harness rows for each metric column; \colorbox{ranksilver}{silver} highlights the global \#2. Definitions of the five meta-categories are given in Appendix~\ref{app:task-inventory}.}
\label{tab:category-breakdown}
\resizebox{\textwidth}{!}{%
\begin{tabular}{@{}llccccccccccccccc@{}}
\toprule
 &  & \multicolumn{3}{c}{Media Production ($n{=}40$)} & \multicolumn{3}{c}{Performance \& Coaching ($n{=}23$)} & \multicolumn{3}{c}{Enterprise \& Compliance ($n{=}23$)} & \multicolumn{3}{c}{Personal \& Education ($n{=}13$)} & \multicolumn{3}{c}{Operations \& Research ($n{=}6$)} \\
\cmidrule(lr){3-5}\cmidrule(lr){6-8}\cmidrule(lr){9-11}\cmidrule(lr){12-14}\cmidrule(lr){15-17}
Backbone & Harness & $B$ & $P$ & \$ & $B$ & $P$ & \$ & $B$ & $P$ & \$ & $B$ & $P$ & \$ & $B$ & $P$ & \$ \\
\midrule
\multirow{2}{*}{Qwen3.5-122B} & \textsc{Terminus-2} & 0.075 & 0.141 & \cellcolor{rankgold}\$0.105 & 0.130 & 0.196 & \$0.107 & 0.130 & 0.132 & \cellcolor{ranksilver}\$0.104 & 0.154 & 0.222 & \$0.082 & 0.000 & 0.104 & \cellcolor{rankgold}\$0.088 \\
 & \textsc{Terminus-KIRA} & 0.075 & 0.195 & \$0.219 & 0.043 & 0.084 & \$0.234 & 0.174 & 0.175 & \$0.204 & 0.154 & 0.205 & \$0.284 & 0.000 & 0.146 & \$0.317 \\
\cmidrule(lr){1-17}
\multirow{2}{*}{GPT-5.2} & \textsc{Terminus-2} & 0.150 & 0.186 & \$0.724 & 0.043 & 0.097 & \$1.048 & 0.087 & 0.104 & \$0.748 & 0.154 & 0.272 & \$0.703 & 0.000 & 0.000 & \$1.075 \\
 & \textsc{Terminus-KIRA} & 0.125 & 0.184 & \$1.581 & 0.000 & 0.051 & \$1.597 & 0.087 & 0.103 & \$1.734 & 0.308 & 0.294 & \$1.865 & 0.167 & 0.167 & \$1.906 \\
\cmidrule(lr){1-17}
\multirow{5}{*}{Gemini-2.5-Flash} & \textsc{Terminus-2} & 0.075 & 0.149 & \underline{\$0.136} & \textbf{0.087} & \textbf{0.156} & \cellcolor{ranksilver}\underline{\$0.099} & 0.000 & 0.008 & \$0.145 & 0.077 & 0.235 & \cellcolor{rankgold}\textbf{\$0.027} & \textbf{0.167} & \underline{0.241} & \cellcolor{ranksilver}\textbf{\$0.123} \\
 & \textsc{Terminus-KIRA} & \underline{0.100} & 0.168 & \$0.252 & 0.000 & 0.046 & \$0.177 & 0.130 & 0.195 & \$0.260 & 0.000 & 0.051 & \$0.206 & 0.000 & 0.104 & \underline{\$0.150} \\
 & \textsc{Terminus-IA} & 0.100 & 0.163 & \$0.316 & \underline{0.043} & 0.068 & \$0.140 & 0.261 & 0.293 & \$0.193 & \underline{0.231} & 0.258 & \$0.191 & 0.000 & 0.140 & \$0.301 \\
 & \textsc{Terminus-IV} & 0.100 & \underline{0.179} & \$0.188 & 0.000 & 0.043 & \$0.191 & \textbf{0.391} & \textbf{0.456} & \underline{\$0.115} & 0.231 & \underline{0.265} & \$0.198 & \underline{0.167} & 0.208 & \$0.372 \\
 & \textsc{Terminus-MM} & \textbf{0.225} & \textbf{0.327} & \cellcolor{ranksilver}\textbf{\$0.131} & 0.043 & \underline{0.130} & \cellcolor{rankgold}\textbf{\$0.096} & \underline{0.391} & \underline{0.424} & \cellcolor{rankgold}\textbf{\$0.050} & \cellcolor{ranksilver}\textbf{0.385} & \textbf{0.355} & \cellcolor{ranksilver}\underline{\$0.047} & 0.000 & \textbf{0.263} & \$0.195 \\
\cmidrule(lr){1-17}
\multirow{5}{*}{Gemini-3.1-Pro} & \textsc{Terminus-2} & 0.125 & 0.172 & \textbf{\$0.734} & 0.087 & 0.157 & \$1.107 & 0.130 & 0.140 & \textbf{\$0.568} & 0.154 & 0.194 & \textbf{\$0.675} & 0.167 & 0.133 & \textbf{\$0.733} \\
 & \textsc{Terminus-KIRA} & 0.075 & 0.127 & \$2.179 & 0.087 & 0.186 & \$1.777 & 0.174 & 0.200 & \$2.031 & 0.077 & 0.154 & \$2.316 & 0.167 & 0.133 & \$1.931 \\
 & \textsc{Terminus-IA} & \cellcolor{ranksilver}\underline{0.250} & 0.343 & \$1.981 & 0.217 & 0.286 & \textbf{\$1.032} & 0.522 & 0.582 & \$1.917 & \underline{0.385} & 0.401 & \$2.040 & \cellcolor{rankgold}\textbf{0.500} & \cellcolor{ranksilver}\underline{0.617} & \$1.562 \\
 & \textsc{Terminus-IV} & 0.225 & \cellcolor{ranksilver}\underline{0.349} & \$1.356 & \cellcolor{ranksilver}\underline{0.217} & \cellcolor{rankgold}\textbf{0.292} & \$1.189 & \cellcolor{rankgold}\textbf{0.565} & \cellcolor{ranksilver}\underline{0.599} & \underline{\$1.242} & \cellcolor{rankgold}\textbf{0.462} & \cellcolor{rankgold}\textbf{0.536} & \$1.379 & 0.333 & \cellcolor{rankgold}\textbf{0.659} & \underline{\$1.109} \\
 & \textsc{Terminus-MM} & \cellcolor{rankgold}\textbf{0.300} & \cellcolor{rankgold}\textbf{0.471} & \underline{\$1.271} & \cellcolor{rankgold}\textbf{0.261} & \cellcolor{ranksilver}\underline{0.290} & \underline{\$1.070} & \cellcolor{ranksilver}\underline{0.565} & \cellcolor{rankgold}\textbf{0.629} & \$1.409 & 0.385 & \cellcolor{ranksilver}\underline{0.431} & \underline{\$1.028} & \cellcolor{ranksilver}\underline{0.500} & 0.615 & \$1.289 \\
\cmidrule(lr){1-17}
Sonnet-4.6 & \textsc{Claude Code} & 0.150 & 0.195 & \$1.640 & 0.043 & 0.087 & \$1.430 & 0.217 & 0.211 & \$2.049 & 0.308 & 0.295 & \$2.178 & 0.167 & 0.167 & \$1.363 \\
\cmidrule(lr){1-17}
GPT-5.2 & \textsc{Codex CLI} & 0.225 & 0.258 & \$6.005 & 0.000 & 0.022 & \$6.444 & 0.174 & 0.257 & \$9.189 & 0.154 & 0.154 & \$7.611 & 0.333 & 0.417 & \$8.101 \\
\bottomrule
\end{tabular}%
}
\end{table*}